\documentclass[9pt]{article}
\usepackage{subcaption}
\usepackage[affil-it]{authblk}

\usepackage{graphicx}
\usepackage{amsmath}
\usepackage{hyperref}
\usepackage{multirow}
\usepackage[utf8]{inputenc}
\usepackage[T1]{fontenc}
\usepackage{makecell}

\providecommand{\keywords}[1]
{
  \small	
  \textbf{\textit{Keywords---}} #1
}

\title{From trust in news to disagreement: is misinformation more controversial?}

\author[a,b,c,d]{D. Ruggiero Lo Sardo}
\author[a,d]{Emanuele Brugnoli}
\author[c]{Enrico Ubaldi}
\author[a,b,c,d]{Vittorio Loreto}
\author[c,d]{Pietro Gravino}
\affil[a]{Sony Computer Science Laboratories - Rome, Joint Initiative CREF-SONY, Centro Ricerche Enrico Fermi, Via Panisperna 89/A, 00184, Rome, Italy}
\affil[b]{Physics Department, Sapienza University of Rome, Piazzale Aldo Moro 2, 00185 Rome, Italy}
\affil[c]{Sony Computer Science Laboratories - Paris, 6, Rue Amyot, 75005, Paris, France}
\affil[d]{Centro Ricerche Enrico Fermi, Via Panisperna 89/A, 00184, Rome, Italy}

\date{}

\begin{document}

\maketitle

\begin{abstract}
The growing prevalence of fruitless disagreement threatens social cohesion and constructive public discourse. While polarised discussions often reflect distrust in the news, the link between disagreement and misinformation remains unclear. In this study, we used data from ``Cartesio'', an online experiment rating the trustworthiness of Italian news articles annotated for reliability by experts, to develop a disagreement metric that accounts for differences in mean trust values. Our findings show that while misinformation is rated as less trustworthy, it is not more controversial. Furthermore, disagreement correlates with increased commenting on Facebook. This suggests that combating misinformation alone may not reduce polarisation. Disagreement focuses more on the divergence of opinions, trust, and their effects on social cohesion. Our study lays the groundwork for unsupervised news analysis and highlights the need for platform design that promotes constructive interactions and reduces divisiveness.

\end{abstract}
\\ 
\vspace{0.2cm}

The press and news media are fundamental parts of democracy. They inform public debate and act as a balance check for the three independent powers of democratic life (legislative, executive, and judiciary). Famously, in 1787, the British member of Parliament Edmund Burke coined the expression ``the fourth estate'', referring to the press~\cite{schultz1998reviving}. More recently, the Italian Constitutional Court referred to the function of journalism as the <<``watchdog'' of democracy>>~\cite{Const2121cane}.

The role of the press in a democracy currently faces significant challenges, particularly due to the decline of trust in media, i.e. the confidence of consumers in the veracity of information provided by traditional sources~\cite{brenan2022americans, Reuters2023}. This decline hampers the press's ability to inform public discourse, which is vital for the democratic process~\cite{gutmann2009democracy}. As a consequence, individuals participating in the public debate increasingly turn to alternative information sources, leading to a heightened risk of becoming spreaders of misinformation, understood as false or inaccurate information disseminated regardless of intent to deceive~\cite{Reuters2023}.

Moreover, the increasing prevalence of misinformation might affect trust in news in a vicious cycle empowered by social networks and generative AI~\cite{kidd2023,spitale2023}. 
Consequently, the World Economic Forum has ranked the spread of misinformation as the most severe global risk in the short term~\cite{WEF2024}.
Prior research has explored various aspects of misinformation spread, including its impact during natural disasters~\cite{muhammed2022disaster}, political crises~\cite{lewandowsky2024}, threats to public health~\cite{wang2019systematic} and psychological well-being~\cite{ecker2022psychological}, and the channels on which it spreads~\cite{Reuters2019}. 

Understanding why people assign different levels of trust to news, especially misinformation, remains an open research question. Studies show that even with similar credibility evaluation criteria, people can interpret the reliability of search engine results differently~\cite{kattenbeck2019understanding}. 
A common explanation for the differing evaluations is partisan affiliation~\cite{peach2024seeing, michael2021relationship}. However, studies have found susceptibility to misinformation is also influenced by cognitive laziness rather than just partisan bias~\cite{pennycook2019lazy, osmundsen2021partisan, muhammed2022disaster}

Evidence suggests a link between decreasing media trust, the rise of misinformation, and increasing polarisation, i.e., the process by which the attitudes or opinions of individuals or groups within a society become increasingly divergent~\cite{baldassarri2008partisans}. For example, perceived political polarisation erodes social trust and hinders collective action~\cite{lee2022social, jones2015declining, layman2006party} while outlets presenting contrary views are perceived as biased~\cite{rapp2016moral, eveland2003impact, fico2004influence}. 
Misinformation feeds into this cycle by presenting biased narratives that align with pre-existing beliefs~\cite{prochaska2023mobilizing, cinelli2020selective}. Further, the erosion of media trust drives individuals towards echo chambers, reinforcing opinion polarisation~\cite{tokita2021polarized}. 


However, the study of polarisation poses significant challenges when investigating populations beyond users of social networks and across all news media topics. Specifically, the analysis of social media platforms tends to over-represent the opinion of vocal minorities~\cite{chenpublic}. Further, when investigating the alignment of opinions not all topic show significant correlations~\cite{baumann2021emergence}.

In this work, we focus on understanding the microscopic structure and emergence of polarisation by investigating disagreement—a measure of differing opinions on specific topics, such as the reliability of news items. 
While distinct concepts, disagreement and opinion polarisation are inherently linked, with disagreement serving as a necessary condition for the emergence of opinion polarisation for which we refer to the definition provided by Koudenburg et al., who describe it as ``the extent to which opinions on an issue are opposed''~\cite{koudenburg2021new}.

By examining the interplay between misinformation and disagreement over trust in news we aim to provide insights into the underlying mechanisms driving societal polarisation dynamics.

Our results are based on the analysis of ``Cartesio'', an experiment developed to investigate the trust in Italian news produced between 09/01/2018 and 12/04/2020~\cite{Cartesio}. The experiment took place online through the ``Cartesio'' portal. There were 5803 participants recruited through online promotion on social networks, participation in science fairs, and word of mouth  (for further information on the population sample, refer to the SI Section 1). The experiment collected comprehensive data on public discourse and trust and the structure of the opinions on the trustworthiness on Italian news. The study gives us a deeper insight into the relationship between news veracity and trust. 

In particular, we find that though misinformation receives lower trust scores, it does not elicit higher disagreement than regular news.
In other words, disagreement in trust scores is not simply driven by the presence of misinformation.
To this aim, we give a rigorous definition of disagreement that encapsulates the degree to which opinions differ. Building on this definition, we introduce a measure of disagreement designed to account for asymmetrical bimodal distributions, which is particularly useful in scenarios where a significant imbalance exists in the size of groups holding different opinions.

Additionally, we investigate how the results of the ``Cartesio'' experiment compare to the engagement received by the same news items as they are posted on Facebook by the news providers. Although the sample of news we could observe is relatively small, we find significant correlation between disagreement and the amount of comments received by the Facebook post.

In the following, we first describe the experiment and the data, the demographic properties of the population sample and the processing procedures. We delve deeper into analysing distributional aspects of individual trust opinions. We then examine the spread of news trust in the totality of the news ecosystem dataset, revealing the overall level of trust in the news of participants and the lack of multiple modes in the distribution. The study also describes the relationship between news veracity and trust. We find that misinformation receives lower trust scores and discuss using mean trust as a discriminatory test for information veracity. Furthermore, we scrutinised the correlation between the trust ratings assigned by our participants and the reactions these news items elicited on Facebook. This detailed analysis aims to identify whether the perceived trustworthiness of news influences public responses on social media, thus offering a comprehensive understanding of trust, misinformation, and the dynamics of social media engagement. 

\section*{The distribution of trust}

\begin{figure}[ht]
  \centering
\includegraphics[width=.5\linewidth]{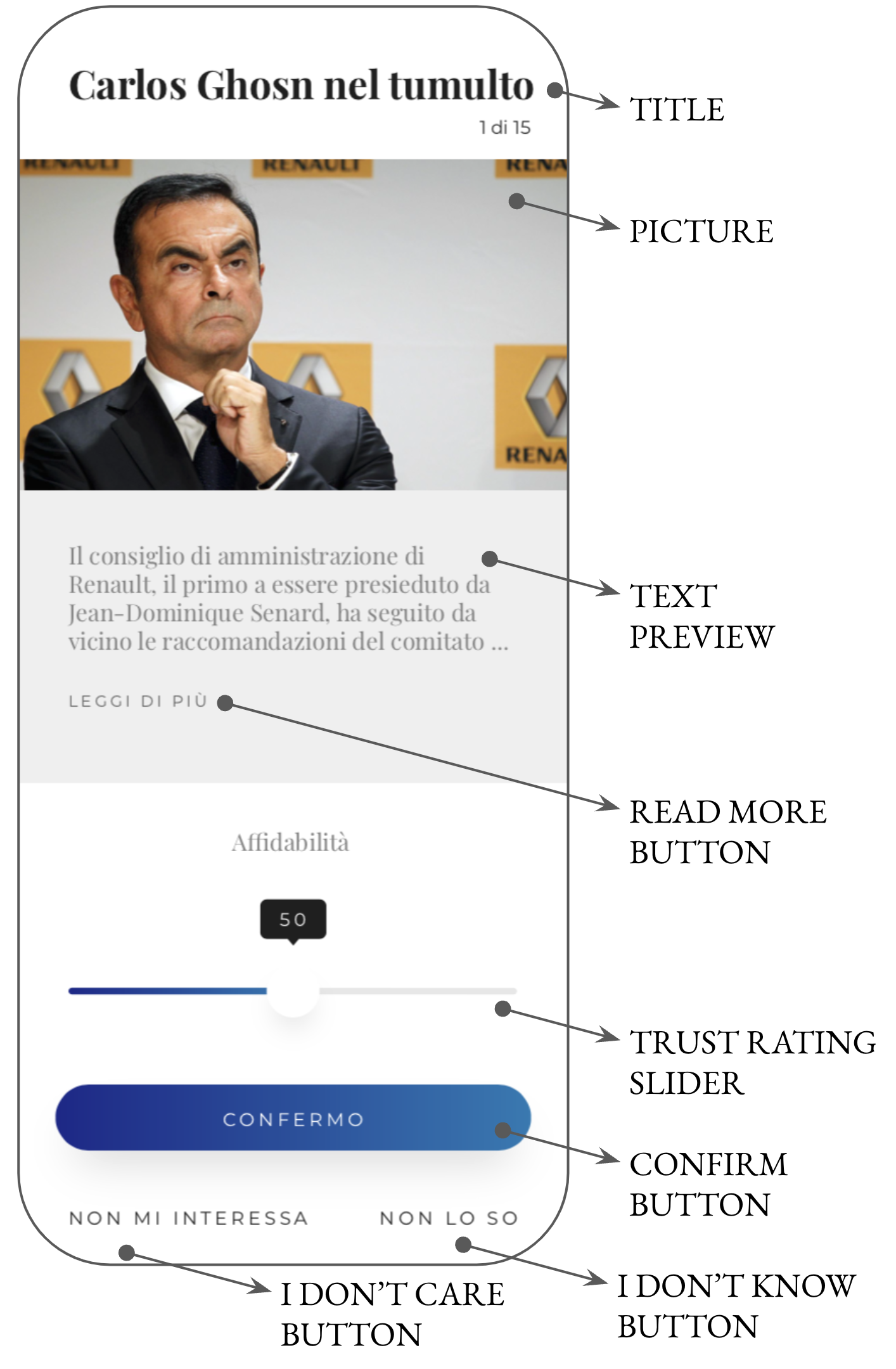}
    \caption{ }
    \label{fig:interface}
\end{figure}

The ``Cartesio'' experiment (see Figure~\ref{fig:interface}) gathered participants' trust evaluation in individual news items belonging to an annotated dataset of newspaper articles (regular, non-misinformation news) enriched with articles that were labelled as misinformation by a trusted flagger. Participants were recruited mainly online through Facebook ads, through word of mouth and at science fairs.

Table~\ref{tab:demo} provides a detailed overview of the demographic characteristics of participants who agreed to share their information. The ``Frequency'' column presents the actual count of participants within each category, while the ``Known $\%$'' column reflects the proportion of these participants among those who provided their demographic data. The ``Italian Demographic $\%$'' column offers a baseline for comparison, sourced from national statistics.

It is important to note that the reference population for our study is not the Italian population at large, but rather the subset of the Italian population that is active on social networks.
\begin{table}[]
    \centering
    \footnotesize
\begin{tabular}{c c c c c}
\hline
\textbf{Characteristics} &  & \textbf{Frequency} & \textbf{Sample $\%$} & \makecell{\textbf{Italian} \\ \textbf{Population $\%$}}\\
\hline
\multirow{9}{*}{\textbf{Age range}} & < 18 & 19  & $3.6$ & 15.1\\
 & 18 - 20 & 33 & $6.3$ & $3.0$\\
 & 21 - 24 & 25  & $4.8$ & $4.0$\\
 & 25 - 34 & 44  & $8.4$ & $10.6$\\
 & 35 - 44 & 88  & $16.9$ & $11.9$\\
 & 45 - 54 & 102  & $19.6$ & $15.5$\\
 & 55 - 64 & 127  & $24.4$ & $15.5$\\
 & > 64 & 64  & $12.3$ & $24.3$\\
 & Unknown & 675 & - & -\\
\hline
\multirow{6}{*}{\textbf{Education}}  & Elementary & 21 & $4.0$ & $11.3$\\
 & Middle-School & 16 & $3.1$ & $33.7$\\
 & High-School & 215 & $41.2$ & $39.0$\\
 & University* & 269 & $51.6$ & $16.0$\\
 & Unknown & 675 & - & -\\
\hline
\multirow{3}{*}{\textbf{Gender}} & Female & 135 & $27.8$ & $51.1$\\
 & Male & 352 & $72.3$ & $48.9$\\
 & Unknown & 675 & - & -\\
\hline
\multirow{21}{*}{\textbf{Regions}} & Abruzzo & 11 & $2.1$ & $2.2\%$\\
 & Basilicata & 7 & $1.3$ & $0.9$\\
 & Calabria & 17 & $3.2$ & $3.1$\\
 & Campania & 46 & $8.8$ & $9.5$\\
 & Emilia-Romagna & 36 & $6.9$ & $7.6$\\
 & Friuli-Venezia Giulia & 13 & $2.5$ & $2.0$ \\
 & Lazio & 93 & $17.8$ & $9.7$\\
 & Liguria & 9 & $1.7$ & $2.6$\\
 & Lombardia & 70 & $13.4$ & $17.0$\\
 & Marche & 18 & $3.5$ & $2.5$\\
 & Molise & 5 & $1.0$ & $0.5$\\
 & Piemonte & 25 & $4.8$ & $7.2$\\
 & Puglia & 23 & $4.4$ & $6.6$\\
 & Sardegna & 9 & $1.7$ & $2.7$ \\
 & Sicilia & 28 & $5.4$ & $8.1$\\
 & Toscana & 30 & $5.8$ & $6.2$\\
 & Trentino-Alto Adige & 8 & $1.5$ & $1.8$\\
 & Umbria & 7 & $1.3$ & $1.4$\\
 & Veneto & 46 & $8.8$ & $8.2$\\
 & Unknown & 695 & - & -\\
\hline
\end{tabular}
    \caption{ }
    \label{tab:demo}
\end{table}

In the age range category, the participant sample shows a substantial under-representation of individuals under 18 years ($3.6\%$ vs. $15.1\%$) and over 64 years ($12.3\%$ vs. $24.3\%$). Conversely, participants aged 45 to 64 are notably over-represented, with the 55-64 age group accounting for $24.4\%$ of the sample, compared to $15.5\%$ in the general population. This over-representation of middle-aged participants may skew the study’s findings toward the perspectives of this age group, potentially limiting the generalisability of the results to the broader population.

The degree category reveals a pronounced over-representation of participants with higher education. While only $16.0\%$ of the Italian population has a university degree, $54.1\%$ of participants in this study fall into this category, with an additional $25.4\%$ holding post-university qualifications. This suggests a bias toward more educated individuals, which could influence the study outcomes given that education level often correlates with various cognitive and behavioural factors.

Gender distribution in the sample shows a significant imbalance, with males comprising $72.3\%$ of participants, compared to $48.9\%$ in the general population. This male over-representation might reflect the study’s thematic appeal or recruitment strategies and could introduce a gender bias in the findings, particularly if gender plays a critical role in the research topic.

Regarding location, while there is a general alignment between the sample and the national population distribution across Italian regions, certain areas like Lazio ($17.8\%$ vs. $9.7\%$) are over-represented, which could affect regional analyses. The substantial portion of participants (695 individuals) whose location data is unknown also introduces uncertainty into these comparisons.

However, some of the observed differences between our sample and the general Italian population may reflect the characteristics of this online-active subset. For instance, the higher representation of regions with large cities, such as Lazio and Lombardia, could indicate a higher concentration of social media users in urban areas, where digital infrastructure and access to technology are more prevalent. Similarly, the skew towards male, middle-aged participants and those with higher education levels may correspond to the demographics of individuals who are active online, as reported by Audiweb for the year 2020~\cite{audiweb2020}.

In order control for the effects of our population sample we compare to ordinary least squares regressions (OLS), one with the raw data and another in which observations are weighted in order to rebalance the demographic representativeness (for futher details on the model see SI Section 3). The two models do not vary significantly in terms of their predictions of the trust ratings, and the effects of the independent variables are substantially compatible. 

Finally, participants engaged through the ``Cartesio'' platform and were assigned to one of four recommender systems (RSs) randomly:  two content-based RSs, a collaborative filtering, and a random algorithm. The analysis of the dynamical effects of the RSs is beyond the scope of this paper, but we report that no effect on the average levels of trust was observed, nor was there any priming effect connected to the time of exposure to misinformation (see SI Section 4 for more details). A snapshot of the interface is reported in Figure~\ref{fig:interface}, and more details can be found in the Materials and Methods section. Building upon the framework introduced by Di Maggio et al.~\cite{dimaggio1996have}, our first insight concerns how individual opinions about the news (i.e., the trust) are distributed within the entire news ecosystem, irrespective of specific news item properties.

\begin{figure}[!ht]
  \centering
    \includegraphics[width=\linewidth]{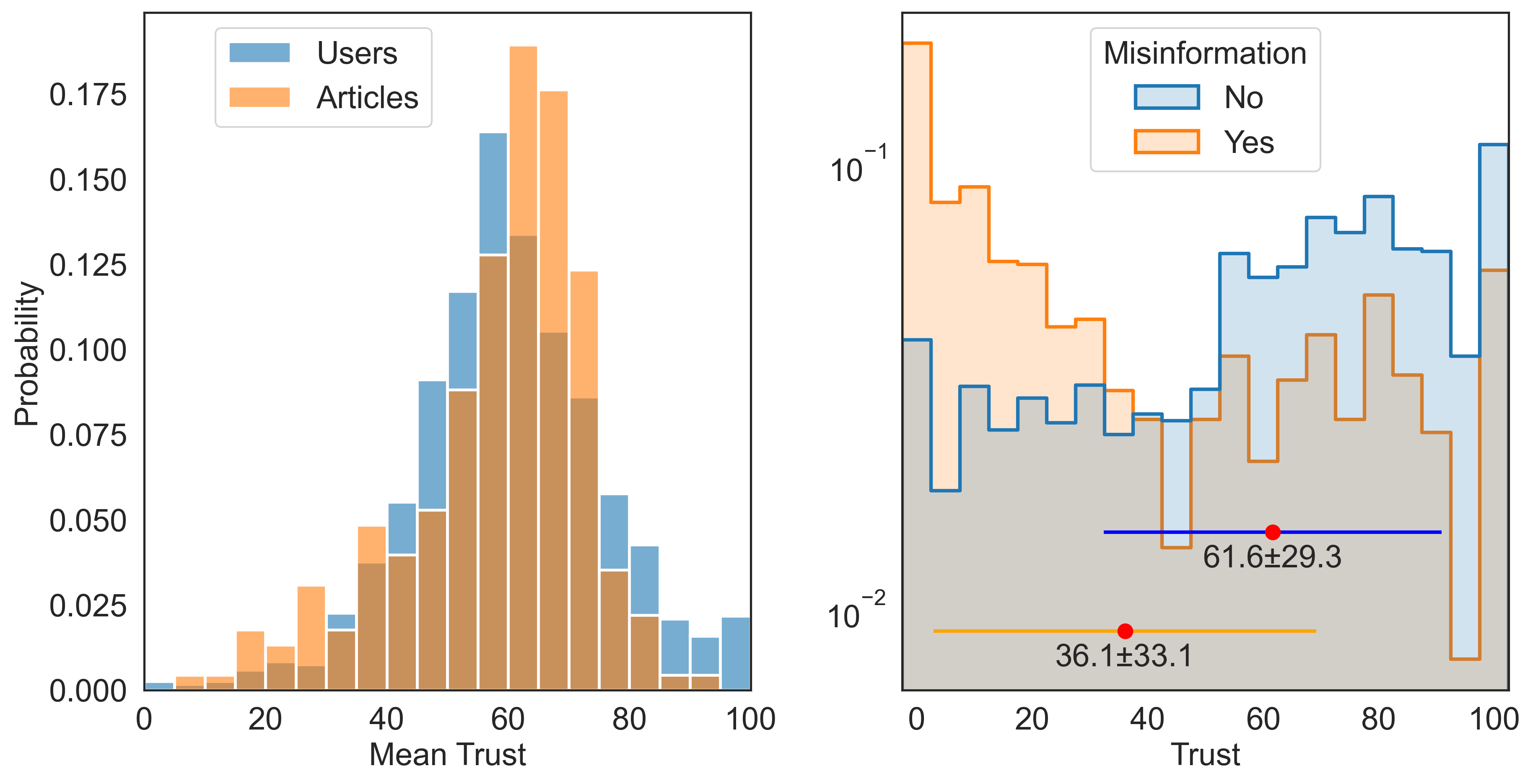}
  \caption{ }
    \label{fig:statistics}
\end{figure}

The global distribution of the trust, irrespective of the participant's identity and news items, covers all possible values, with a higher density of ratings above the midpoint of the rating space (which is from 0-no trust-to 100 for maximum trust), an expectation value of $61.6$, and a standard deviation of $29.3$. The extremal values for the rating are both highly probable. Figure~\ref{fig:statistics} reports the distribution of users' mean trust, i.e., the distribution of the average ratings given by individual users. This distribution demonstrates a well-defined typical behaviour, i.e. bell-shaped curves with an average of $59.8$ and a standard deviation of $15.6$, so no group structure appears at this level of analysis. The same seems true for the distribution of articles' mean trust, with an average of $57.7$ and a standard deviation of $15.1$ (for further information on the distribution of trust see SI Section 3).
Focusing on misinformation articles (i.e., the pieces of news in the dataset labelled as misinformation), we observe a marked difference in trust distribution, with significantly lower trust scores (see the orange share in the bottom panel of Figure~\ref{fig:statistics} and SI Section 3.1 for an OLS modellisation of trust). This disparity in trust values highlights the effectiveness of participants' assessments in differentiating between the two categories. When employing the mean trust value of a news item as a binary classification test (with the negative meaning that the article is misinformation and a conversely for the positive prediction), we achieve an area under the receiver operating characteristic curve (AUC) of 0.88. This value is the probability that a randomly selected misinformation news item receives a lower mean trust score than a randomly selected regular news item. 
However, misinformation is not always distrusted. We can see how also the higher part of the trust score range is populated. Conversely, we also observe supposedly reliable news being rated lower on the scale. This evidence suggests the existence of a certain degree of discord, i.e. a property of certain news items to generate an heterogeneous trust distribution that covers the whole trust range.

\begin{figure}[ht]
    \centering
    \includegraphics[width=\columnwidth]{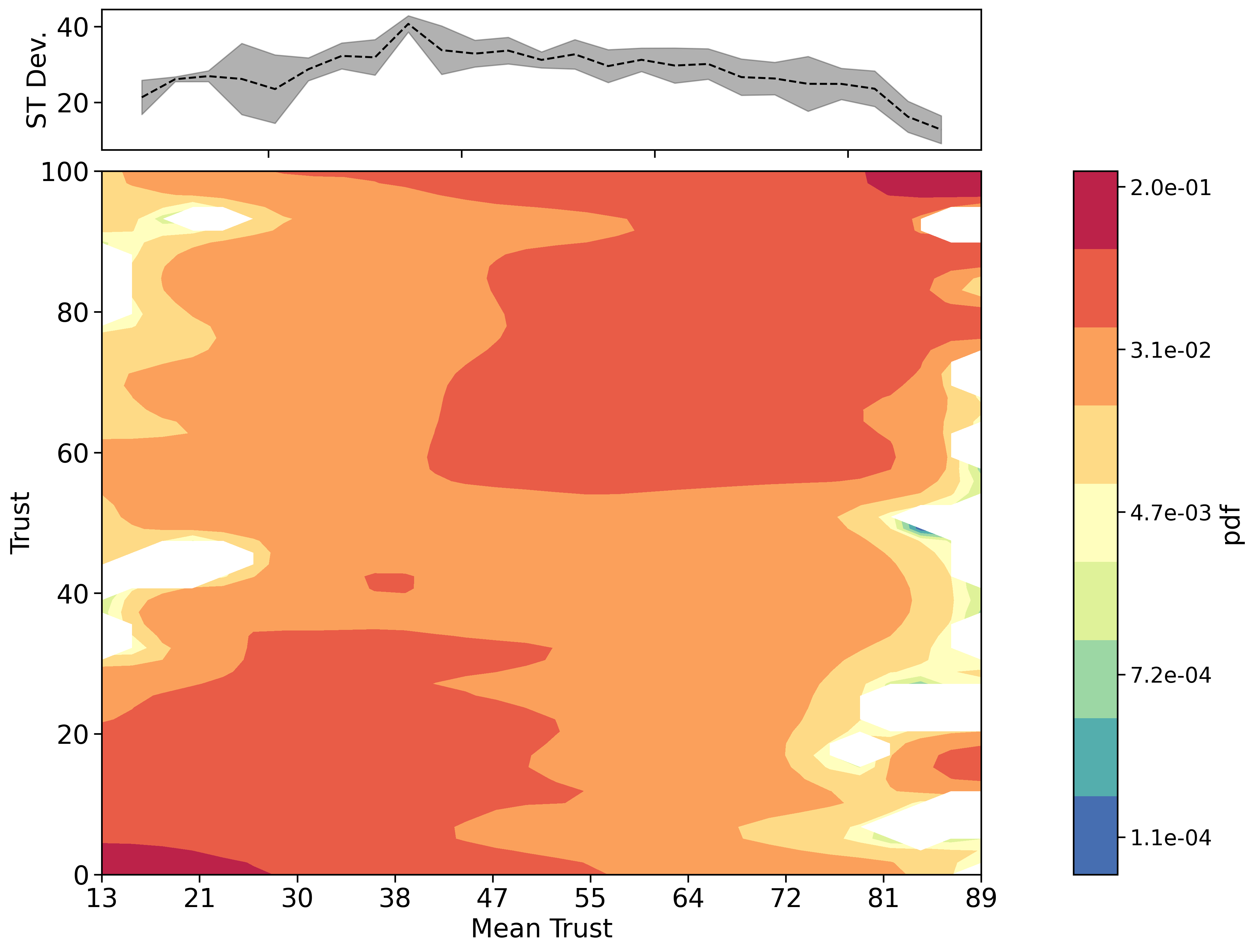}
    \caption{ }
    \label{fig:1}
\end{figure}

To further investigate the discord and its effect on trust distribution, we examined how the latter depends on the mean trust obtained by different news items. We grouped news items that received comparable mean trust and looked at each group's trust distribution, normalising each distribution to be comparable. As depicted in Figure~\ref{fig:1}, the x-axis represents the mean trust of news items, while the y-axis displays individual trust ratings. The heatmap illustrates the density of trust ratings that contributes to evaluating news items within a particular mean trust range.

Our findings reveal unimodal distributions on news items with average evaluations at the extremes of the reliability scale, suggesting a general agreement among participants. Conversely, we observed a distinct lack of coherence, i.e. a unified whole, for news items in the middle of the mean trust spectrum. This evidence suggests that a consensus on reliability was absent for this kind of news.

Looking at this difference in dispersion through the lens of standard deviation is not very informative. The dispersion distribution has its lowest value at the extremes of the mean trust spectrum and the highest in the central part. However, this was to be expected given the known relation between mean and standard deviation in bounded distribution ~\cite{popoviciu1935equations}. A correction to this effect will be discussed in the following section.

\section*{Disagreement and Misinformation}

Expanding upon earlier formulations of distributional polarisation~\cite{dimaggio1996have}, we introduce a novel metric called \emph{Disagreement}, which emphasises the dispersion principle while accounting for the limitations of standard deviation in bounded distributions~\cite{popoviciu1935equations}. In other words, while standard deviation inherently goes to zero for distributions with averages close to the bounds' of the distribution support, \emph{Disagreement} allows for comparisons between dispersions of bounded distributions with different mean values. This measure is particularly relevant when two or more groups hold divergent opinions, i.e. different modes of the trust distribution coexist, with one group (or mode) being significantly larger than the other. In such cases, the standard deviation may be constrained by this asymmetry, even though the differences between opinion groups are of social importance.
We correct the dispersion by normalising with respect to the maximal dispersion possible given a certain mean trust on a finite interval. 
In practice, \emph{Disagreement} $D$ is defined as:
\begin{equation*}
    D = \frac{\sigma}{\sqrt{(\mu-a)(b-\mu)}}
\end{equation*}
where $\sigma$ is the standard deviation around the mean, $\mu$ is the mean of the distribution, and $a=0$ and $b=100$ are, respectively, the minimal and maximal values of the interval. \emph{Disagreement} is, by definition, bounded between $0$ and $1$.

\begin{figure}[!ht]
    \centering
    \includegraphics[width=\columnwidth]{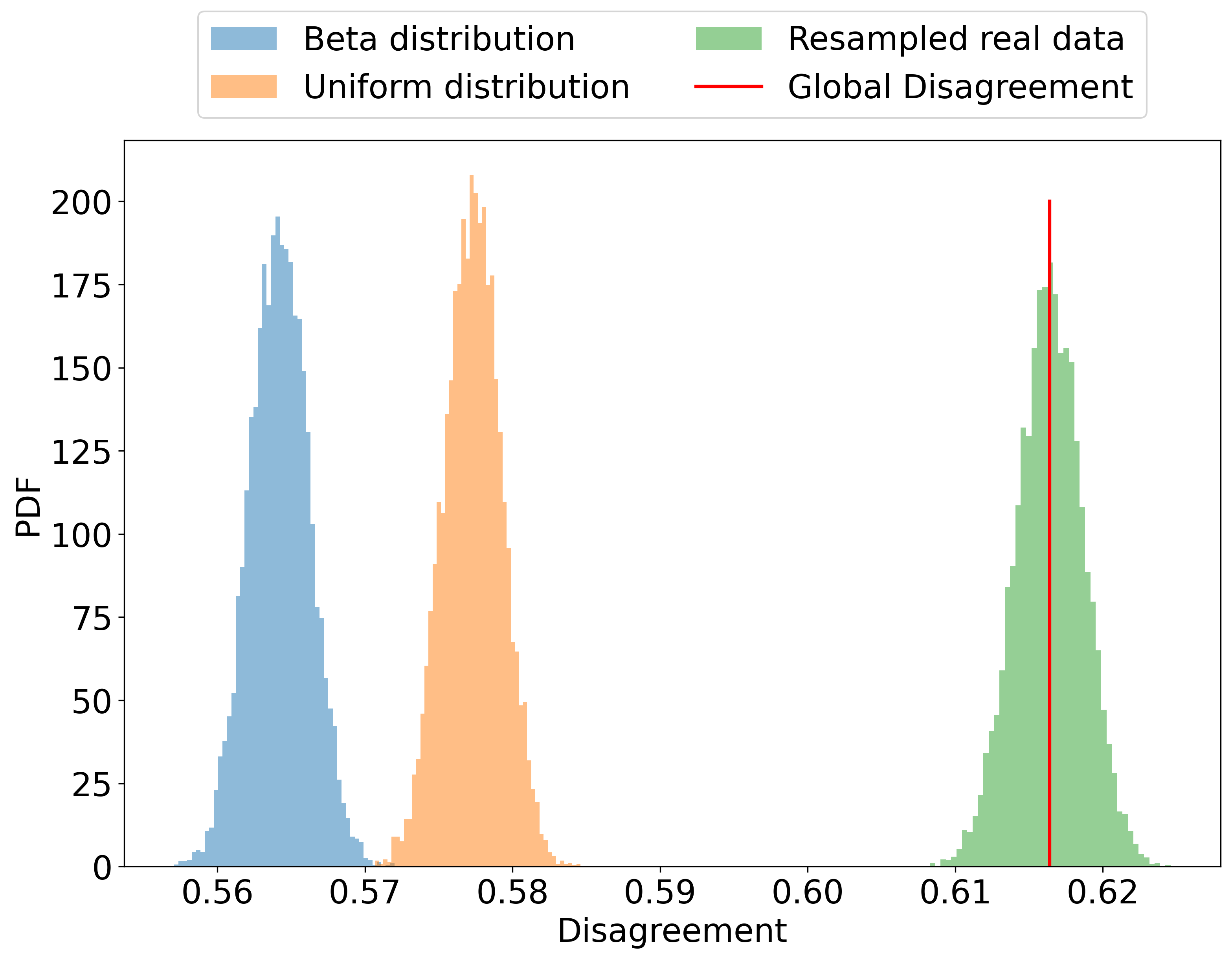}
    \caption{ }
    \label{fig:synthEx}
\end{figure}

We analyse the \emph{Disagreement} in the dataset concerning synthetic distributions to contextualise the current state of opinions on trust in the media (see Figure~\ref{fig:synthEx} and Materials and Methods). Our findings indicate that \emph{Disagreement} in the global trust distribution is significantly larger than the ones found when drawing trust scores from both a uniform and a best-fit beta distribution, signalling a degree of polarisation of the trust on the individual news items that cannot be attained with such null models.

\begin{figure}[ht]
    \centering
    \includegraphics[width=\columnwidth]{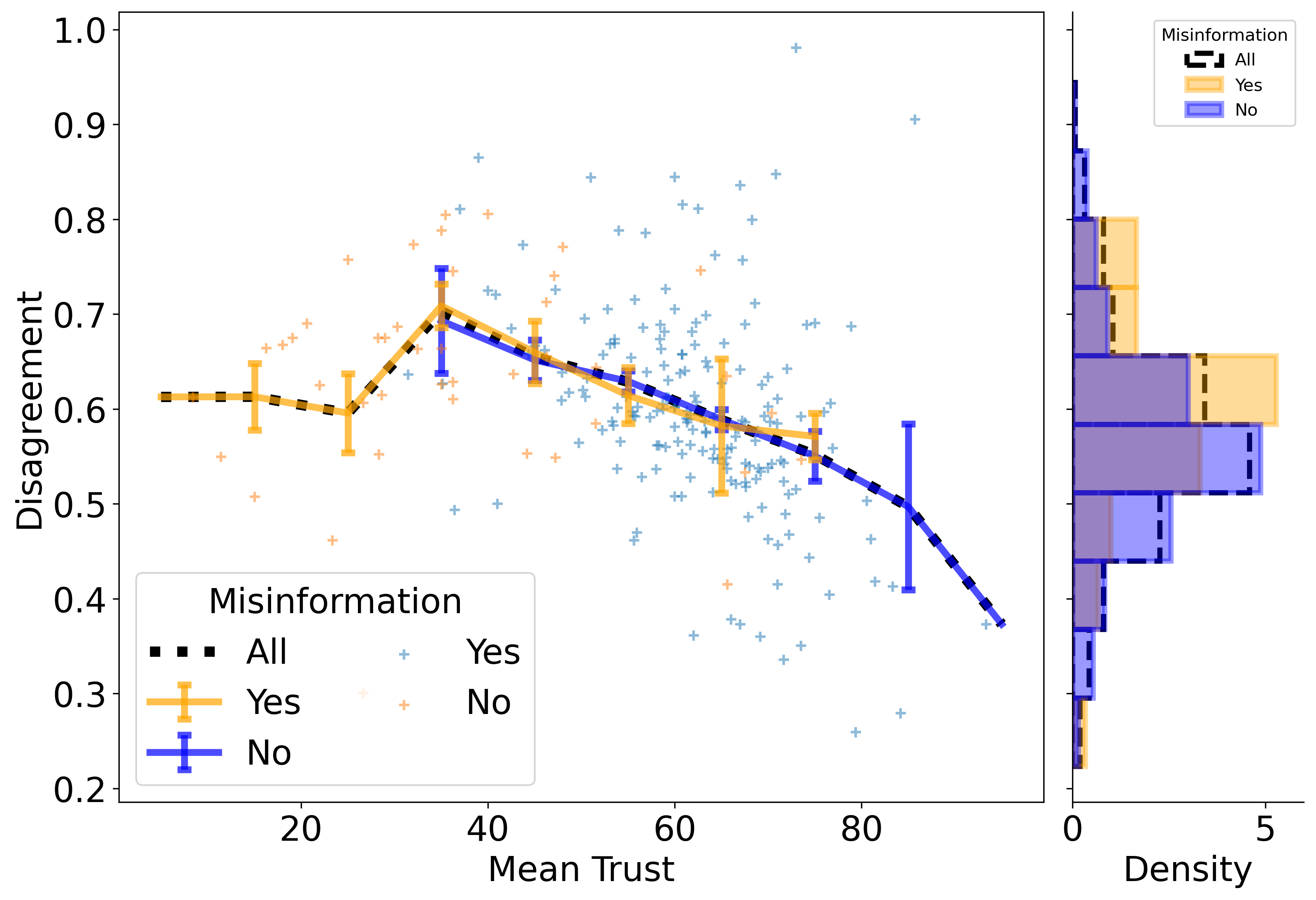}
     \caption{ }
    \label{fig:3}
\end{figure}
We examined the \emph{Disagreement} of individual news items in more detail, analysing the relation with the mean trust (see Figure~\ref{fig:3}). Our results reveal that \emph{Disagreement} is lower for news items with high mean trust compared to the overall distribution. Furthermore, we observed that maximum \emph{Disagreement} occurs for news items located just below the midpoint of the trust range, corroborating the visual representation in Figure~\ref{fig:1}. In Figure~\ref{fig:3}, we also report a comparison between the non-trustworthy news items and other news, and the two curves are not distinguishable when accounting for variance. In other words, it seems that non-trustworthy news generates the same levels of \emph{Disagreement} as the others, suggesting that news veracity is not directly linked to polarisation. Further, using a linear model to describe the overall behaviour of ratings ($D \sim a \times \textit{trust scores}+b$), we observe that the residual distribution is the same for a model fitted to the misinformation data-points and a model fitted to regular news (Kolmogorov-Smirnov statistic of 0.175, p-value of 0.48).



\section*{Trust and Online Reactions}
We further extended our analysis by pairing the news items rated by the participants in our experiment with the corresponding Facebook posts shared by the official profiles of the news sources (see the Material and Methods for details on the matching procedure). Our aim was to investigate the relationship between the participants' trust rating distribution and the reactions elicited on Facebook. This pairing allowed us to associate the trustworthiness scores assigned by the participants with the nature and volume of engagement these news items generated on Facebook, including likes, emoticon-like reactions, shares, and comments. By drawing comparisons between the two data sets, we aimed to decipher if there was a correlation between the perceived trustworthiness of a news item and the public reaction it garners on social networks. This analysis provides insight into the intricate dynamics of trust, misinformation, and social media engagement. 

\begin{figure}[!ht]
    \centering
    \includegraphics[width=\columnwidth]{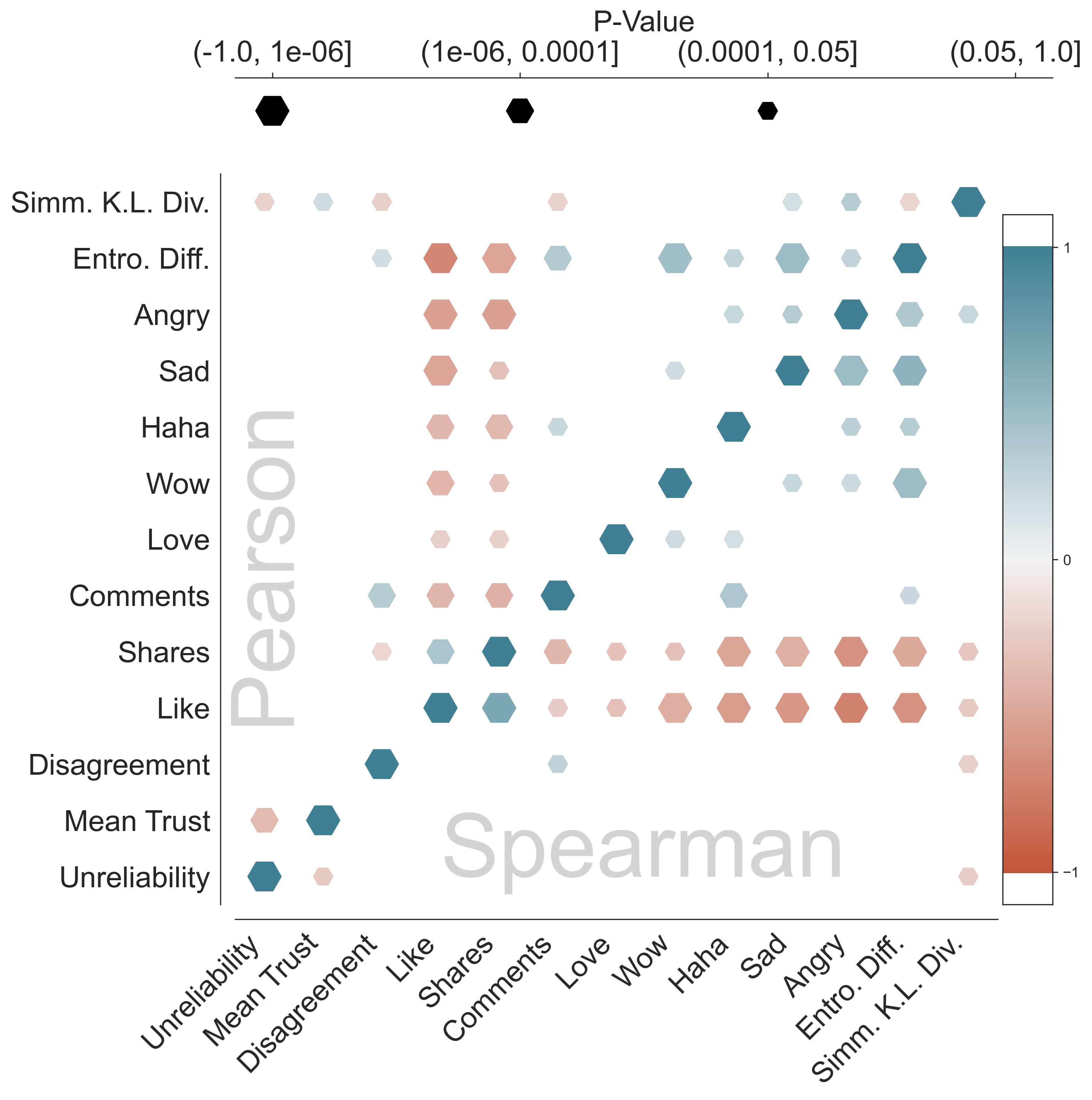}
    \caption{ }
    \label{fig:corrplot}
\end{figure}

The first step in this analysis is to understand the relationship between raw interaction metrics and popularity. As a matter of fact, a first analysis of these metrics displays how all of these measures are strongly correlated (consistently with previous findings~\cite{freeman2020measuring}). Due to the platform mechanism, the metrics are strongly influenced by the popularity of the source profile. For this reason, we normalise metrics with respect to the source and post-type. Additionally, we normalise the metric so that the standardised values sum up to 1, to remove the effect of popularity of individual posts. In fact, a popular post will receive, on average, a greater amount of all sorts of reactions, causing a positive correlation among them that would hide the signal. Finally, we measured the correlations between the Facebook normalised metrics and Cartesio metrics to study the relations. The results are reported in Figure~\ref{fig:corrplot}.

The strongest positive correlations among Facebook metrics are between reactions with close meanings, as \emph{angry} and \emph{sad}, that share a negative sentiment, or as \emph{shares} and \emph{like}, that share a positive one. Similarly, these meanings emerge as a contrast for negative correlations (like \emph{sad} and \emph{like}). It is worth noting that \emph{comments} do not have strong correlations, and the weak ones tend to be in an ambiguous position between positive sentiment (e.g., positive correlation with \emph{like}) and negative sentiment (positive correlation with \emph{ahah}).

Regarding the relationship between Facebook metrics and Cartesio metrics, we find that news with higher \emph{Disagreement} displays more comments (Spearmans' $\rho = 0.4$). This evidence suggests that articles generating \emph{Disagreement} generate, as a side effect, more extended discussions without a clear sentiment, which might be the statistical signature on social networks of polarising content. Furthermore, unreliability and mean trust do not have any meaningful correlation with interaction metrics, suggesting the lack of a statistical signature of misinformation in the social media interaction space.
We measure differences in the Shannon entropy of the distribution of reactions between the overall distribution of reactions and those observed for each news item. We believe entropy to be a good proxy for diversity in ratings since it is maximal when each reaction is equally probable. However, we find no significant correlation with the difference in entropy. We further checked for the effects of the symmetrised Kullback–Leibler divergence between the overall distribution of reactions and those received by each news item. This measure captures differences in the distribution of reactions, detecting if they are likely to originate from the same underlying phenomena. Here, we observe a significant correlation with both \emph{Disagreement} and Unreliability, evidence that there is a signal in the reaction distribution regarding these two measures.

\section*{Conclusions}

The prevalence of sterile disagreement poses a significant threat to social cohesion and meaningful public discussions. This disagreement is frequently associated with spreading misinformation, but the specific relationship between misinformation and its role in driving disagreement and opinion polarisation remains unclear.

Throughout this study, we have analysed how trust in news is structured and the defining characteristics of misinformation within this framework. By conducting an experiment in which participants were asked to rate Italian news articles, we observe landscape where misinformation, despite receiving lower trust ratings, did not seem to cause more controversial outcomes.

Our developed metric for disagreement, which accounted for both predominant opinions and distinct opinion groups, provided valuable insights into the nature of contentious discourse. Using both artificial trust distributions and our experimental data, we were able to outline a comprehensive profile of \emph{Disagreement} at both the collective and individual news item levels.

We contend that this finding indicates that labelling misinformation might not be the best strategy for reducing tensions in social dialogue. While labelling misinformation is an important activity for the health of the media landscape, by making it possible to score the reliability of sources and enabling users to source content, the perception that this is often a partisan activity can foster more significant division~\cite{reinero2023partisans}. Reporting \emph{Disagreement} of news articles or including it in feed algorithms can be a pathway to more constructive social dialogue. 

Significantly, our findings underscored the complex relationship between the reliability of news and the patterns of its social media engagement. We found \emph{Disagreement} was associated with a higher likelihood of commenting, revealing the paradoxical ways contentious news can circulate and stimulate discussion within digital spaces. Although stimulating online discussion in the news can be seen as a net positive for the collective, there is evidence that longer discussions lead to higher levels of toxicity~\cite{cinelli2021dynamics}. If it were to hold at scale, this observation would constitute a potential warning for the kind of material that could be more divisive and, therefore, problematic for social dialogue. Platforms that prioritise content that triggers engagement, such as comments, can inflate the salience of controversial topics by increasing their visibility. This, in turn, can lead to increased stress and anxiety in users who will try to avoid negative interactions. It has also been shown that this causal chain can explain the birth of echo chambers~\cite{powers2019shouting, brugnoli2019recursive}.

The work presented here can be improved in many ways. The primary concern is the representativity of the participant set. The study can be clearly improved by selecting a sample that is more balanced in terms of gender and age. By replicating the experiment on a larger population and different countries, we can test the validity of our conclusion and the universality of phenomenology. Further, the potential absence of certain viewpoints could affect the detection of ideological divides in the trust space, while the scarcity of data in some areas, particularly regarding specific topics, limits the study of dependencies of trust within specific topics and how \emph{Disagreement}, a `microscopic' feature, is linked to the political polarisation of the society, which can be considered a structural property of the whole system.

This research lays a solid foundation for unsupervised news item analysis that listens to and leverages users' opinions. While fact-checking and generally high professional standard journalism are crucial against misinformation, more is needed to fight polarisation. The proposed approach focuses on exploring the diversity of opinions and their impacts on social cohesion. It highlights the need for innovative strategies that can integrate our understanding of \emph{Disagreement} and misinformation into the design of news distribution platforms and social media systems. By doing so, we can enhance online interactions, foster a more informed public, and pave the way for the evolution of a more reliable digital news ecosystem. This is a critical step towards mitigating disagreement, bolstering social harmony, and reinvigorating constructive discourse in our society.

\section*{Abbreviations}
\begin{itemize}
    \item \textbf{AGCOM}: Autorità per le Garanzie nelle Comunicazioni (Italian Communications Regulatory Authority)
    \item \textbf{AUC}: Area Under the Receiver Operating Characteristic Curve
    \item \textbf{CSL}: Computer Science Laboratories
    \item \textbf{EU}: European Union
    \item \textbf{GDPR}: General Data Protection Regulation
    \item \textbf{OLS}: Ordinary Least Squares
    \item \textbf{RS}: Recommender System
    \item \textbf{SI}: Supplementary Information
\end{itemize}

\section*{Materials and Methods}

\subsection*{The ``Cartesio'' Experiment}
To measure citizens' level of trust in news articles, we created an online experiment that engaged citizens in an open, participatory experience. Recruitment was done through online promotion, particularly on social media, and dedicated articles in the press. Participants could join the project through an online platform or by downloading the app. The experience began with anonymous registration, where participants could also share demographic information: age, gender, level of education, and location (at a regional level). All users signing in to the experience agreed to the use of information on their use of the platform through an informed consent form. Then, after topic selection, the participants were prompted with news articles they could rate according to their subjective level of trust between 0 and 100. The prompt consisted of a title, an optional subtitle, an image, and a preview of the text that could be expanded (see Figure~\ref{fig:interface}). 
Participants were instructed to provide reliability scores in the range $[0-100]$ by increments of 5 or to respond: ``not interested'' or ``don't know''. These analyses do not include answers that did not fall in the numeric interval. After the evaluation, another article was prompted, and so on. A recommender system chose the following article. In fact, each participant was randomly assigned to a recommender system at the start of the experience. Different basic versions of state-of-the-art algorithms were implemented~\cite{roy2022systematic}. E.g. a content-based recommender exploiting article topic detection, a collaborative filtering recommender based on the similarity of user interaction patterns, and a random algorithm.
After providing at least 10 evaluations, the participant could continue the experience, also gaining access to more information about trust in the news and personalised statistics on the experience named ``trust profile'', followed by the overall average daily trust and participation statistics. 

The data-gathering campaign started in April 2020 and finished in November 2021. There were a total of 5803 participants who contributed 28202 ratings. For the analysis presented in this paper, we kept new items that received 4 ratings from participants that rated at least 4 articles, reducing the dataset to 227 news items, 1196 users, and 17175 ratings.

\subsection*{The news dataset}

The news items dataset was released for the purpose of the experiment by AGCOM, the Italian authority for guarantees in communication. The dataset was selected from the general news production of the main Italian newspapers plus some questionable sources, sampling, when possible, at least one article per topic and per source each week for a period spanning 2018-01-08 to 2020-04-12. This approach ensured a certain level of representativeness of the general social dialogue in the same period. The topics were "Istruzione" (Education), "Salute" (Health), "Sicurezza e giustizia" (Public safety and justice), "Scienza e tecnologia" (Science and Technology), "Immigrazione" (Immigration), "Diritti della persona" (Civil rights), "Economia e lavoro" (Economics and Work), "Istituzioni e political" (Institution and politics), "Ambiente" (Environment), "Infrastrutture e mobilità" (Infrastructure and mobility). The only general interest topic that was not considered is `Sport'.
The original dataset comprised 6955 news items classified by source, date, and reliability label. This label was assigned by AGCOM, leveraging the work of professional fact-checkers. The content of each article was structured as follows: title, subtitle, content, and category label.

\subsection*{Synthetic Data \emph{Disagreement}}
The comparison of the global \emph{Disagreement} to synthetic data reported in Figure~\ref{fig:synthEx} was obtained by testing 3 different sampling strategies: resampling the observed distribution by fixing the probabilities of each review value irrespective of news category or user type, sampling the best fit $\beta$-distribution, and sampling the uniform distribution.
In all three sampling strategies, the dataset size (17175) was kept fixed as the possible values for the rating. The best fit $\beta$-distribution was computed using the Maximum Likelihood Estimation method with respect to the observed data for the values of $\alpha$ and $\beta$ that uniquely identify the distribution. Each distribution was sampled 10,000 times to investigate the variance of the \emph{Disagreement} given the sampling strategy.

\subsection*{Facebook dataset and matching procedure}
The Facebook dataset was obtained by downloading the production of posts on the public pages corresponding to the sources in the news dataset from 2018-01-08 to 2020-04-12 through CrowdTangle (a public insights tool owned and operated by Facebook). The individual articles from the news dataset were matched to the ones in the Facebook dataset.
The matching procedure was based on a few basic principles. First, the public profiles of the news outlets present in our news database were identified. Then, we matched individual news items by looking at all posts published in a one-week time window around the publication date, and we selected posts with high textual overlap (more than 60\% of the length of the shorter text). This subset was then manually validated by the authors. We identified $129$ posts matching $76$ news (many posts can correspond to the same news item due to updates or reposts). To better comprehend the result of our procedure, we compared the popularity distribution of the matched set with the general popularity distribution of all posts produced in the same period ($2375640$ elements). The matched set was more popular than the general posts production in terms of the number of `like' and `share': median values were almost double ($5$ versus $25$ for 'share' and $18$ versus $38$ for `like'). This evidence suggests that the matched set samples the most engaging part of the social discourse. 

\section*{Declarations}

\subsection*{Ethics Statement \& consent to participate}
In this study, 5803 participants provided their freely-given, informed consent to participate and for their data to be used exclusively for research purposes. The procedures followed complied with the ethics requirements adopted in Italy and France. Data processing was in accordance with the General Data Protection Regulation (Regulation EU 2016/679) and the Italian Law 675 of 31 December 1996 - Protection of Persons and Other Subjects Regarding the Processing of Personal Data. The dataset of news items was sourced from the Communications Regulatory Authority (AGCOM), and CrowdTangle data were processed in accordance with its terms of service.

Since Sony CSL is not a public research centre it is not required to have an ethics committee for the evaluation of research projects. Despite this, ethical considerations and guidelines have been meticulously adhered to throughout the project's duration. All research activities were conducted with a continuous commitment to ethical integrity, ensuring that the principles of ethical research were upheld at every stage. As the project has reached its conclusion and the findings are being presented, we affirm that the research was executed in compliance with the ethical standards of both the academic community and the relevant legal requirements.
At inception and during data-gathering phases, none of the authors was actively employed by public institutes that require ethical committees. This project has now concluded, and it is pertinent to note that the ethics committee at Sapienza University of Rome did not formally evaluate the project, as the approval process for ethical compliance at our institution is designed to be conducted ex-ante. 

In the course of our study, which involved the development and utilisation of an application by AGCOM and Sony Computer Science Labs (CSL), we adhered to stringent informed consent and data privacy protocols. Participants in the study were provided with a clear and concise explanation of the research objectives, particularly our focus on understanding user perceptions of the reliability of freely available online news and their views on various socio-economic phenomena.

Prior to participating in the study, individuals were required to provide informed consent, explicitly agreeing to the terms of data use and collection as outlined in the application. We ensured that the consent process was transparent and user-friendly, with participants acknowledging their understanding of the research's scope and the use of their data.

Key points in our ethical approach included:

\begin{enumerate}
    \item Use of Personal Data: The only personal data collected was an email address, which served as the username for the participant's account. This measure was designed to minimise the collection of personal information, aligning with our commitment to data minimisation.
    \item Privacy Assurance: Our data handling processes were in strict compliance with the privacy policies of AGCOM and Sony CSL, ensuring the highest standards of data protection and confidentiality. These policies are publicly available for review.
    \item Optional Demographic Data: Additional personal data such as gender, age, and residence were collected solely for statistical aggregation purposes and their provision was entirely optional.
    \item Data Management: The management of personal data was jointly overseen by Sony and AGCOM, in accordance with relevant privacy regulations and guidelines.
    \item Data Anonymisation and Deletion: At the conclusion of the research, all email addresses were deleted and the data anonymised, ensuring no personal identifiers were retained.
\end{enumerate}

\subsection*{Consent for publication}
Not applicable.

\subsection*{Availability of data and material}
The Cartesio dataset used in this study, consisting of data from news articles and the reviews provided by study participants, can be made available upon reasonable request. Due to the nature of the data, which includes confidential reviews given by study participants in the Cartesio experiment, there are restrictions on its availability. These reviews were gathered under the agreement that they would remain confidential and be used solely for academic research. Therefore, access to this dataset will be granted exclusively for legitimate academic and research purposes. Researchers interested in accessing this dataset should contact the corresponding authors. Each request will be considered in line with the ethical agreements made with the participants and the governing data protection and privacy laws.

We are unable to share the raw data obtained from CrowdTangle\footnote{https://help.crowdtangle.com/en/articles/4558716-understanding-and-citing-crowdtangle-data} but any researcher can gain access to CrowdTangle platform upon request as of 06/2024.

\subsection*{Competing Interest Statement}
The authors declare no competing interests. This work was conducted independently and impartially. There have been no financial, personal, or professional influences that could be construed as a potential conflict of interest in the design, execution, interpretation, or reporting of this research.

\subsection*{Funding}
No external funding was received.

\subsection*{Authors' contributions}

\textbf{D.R. Lo Sardo}: Conceptualization, Methodology, Software, Validation, Formal analysis, Investigation, Data Curation, Writing - Original Draft, Visualization. \textbf{D.R. Lo Sardo} led the conceptual development of the study, designed and implemented the computational models, and was primarily responsible for the original draft of the manuscript. He also played a key role in data analysis and visualization.

\textbf{E. Brugnoli}: Software, Validation, Investigation, Data Curation, Writing - Review \& Editing. \textbf{E. Brugnoli} contributed to the analysis of online reactions, the data gathering and matching and in the validation of results. He was instrumental in acquiring the data and provided critical feedback and editing of the manuscript. 

\textbf{E. Ubaldi}: Software, Validation, Investigation, Data Curation, Writing - Review \& Editing. \textbf{E. Ubaldi} contributed to the analysis of online article content, the data gathering and labeling. He was instrumental in acquiring the data and provided critical feedback and editing of the manuscript. 

\textbf{P. Gravino}: Conceptualization, Methodology, Formal analysis, Data Curation, Writing - Review \& Editing, Supervision. \textbf{P. Gravino} initially conceptualized the study and played a major role in managing the project partners. He was involved in data collection and curation, and provided significant input in reviewing and editing the manuscript. \textbf{P. Gravino} also supervised the project and provided guidance on the overall research direction.

\textbf{V. Loreto}: Conceptualization, Resources, Writing - Review \& Editing, Supervision, Project Administration, Funding Acquisition. \textbf{V. Loreto} was involved in the initial conceptualization of the study and played a major role in securing funding. He oversaw the project administration and provided substantial inputs in manuscript review and supervision.

\section*{Aknowledgements}
The authors wish to express their sincere gratitude to all those who contributed to this research and made it possible. This study was a product of the collaborative efforts of multiple organisations. Our deep appreciation goes to Alessia Leonardi, Maria Luce Mariniello and Gennaro Ragucci from Autorità per le Garanzie nelle Comunicazioni (AGCOM) for their support and, in particular, for providing the news items and labels that made the Cartesio project possible. Special recognition is due to the citizens who actively participated in the project, giving subjective evaluations of news items and contributing significantly to understanding how information is perceived and which topics are deemed as priorities. The anonymity and privacy of all participants were upheld throughout the process. We would also like to thank Riccardo Corradi, Lone, Interact and Lorenzo Stoduti for their work developing and maintaining the Cartesio platform.

\section*{Figure Legends}

\textbf{Figure \ref{fig:interface}}: An example page of the Cartesio experience. Participants were shown the news article and could click the ``LEGGI DI PIÙ'' button to expand the text, use the slider to rate their trust in the article, or use the ``NON MI INTERESSA'' and ``NON LO SO'' buttons that mean respectively ``I am not interested'' and ``I don't know''. The trust rate can be assigned using the slider spanning between 0 (no trust) to 100 (full trust).

\textbf{Figure \ref{fig:statistics}}: \emph{Left.} The average trust scores distribution found when aggregating the overall ratings by user (blue) and article (orange). \emph{Right.} The distribution of trust in the news labelled as misinformation (orange) and not (blue). The histograms display the probability of each trust bin (size=5) for items labelled misinformation and not. The top horizontal line displays the mean and standard deviation of the trust rating given to non-misinformation news items. In contrast, the bottom one displays the mean and standard deviation of the rating given to misinformation.

\textbf{Figure \ref{fig:1}}: \textbf{Top.} The trust score's standard deviation as computed on news belonging to the different mean trust value bins. The line indicates the mean standard deviation, while the shaded area is the root of the variance. The parabolic shape is to be expected in a bounded interval. \textbf{Bottom.} Trust distribution as a function of the average rating obtained by the rated news article. The horizontal axis displays the mean trust received by news items. The vertical axis is the value of individual trust ratings. The colour indicates the density of trust scores received by news items with a given mean trust.

\textbf{Figure \ref{fig:synthEx}}: The distributions of \emph{Disagreement}. In red we report the actual, overall value from participant evaluations of news. We compare it to synthetic distributions obtained by sampling $10,000$ times the same number of votes as the actual dataset (17,175) from a uniform distribution (in orange), a $\beta$-distribution with mean and variance equal to those observed in the dataset (in blue), and the empirical distribution of the experiment, represented in Figure~\ref{fig:statistics} (in green).

\textbf{Figure \ref{fig:3}}: The relationship between mean trust and user \emph{Disagreement} for misinformation and other news items. Each point represents a news item in the space defined by mean trust (horizontal axis) and \emph{Disagreement} score (vertical axis). Blue and orange lines depict average \emph{Disagreement} trends for regular news and misinformation, with error bars indicating standard deviation. Histograms on the right show \emph{Disagreement} distribution for all news items and misinformation labels.

\textbf{Figure \ref{fig:corrplot}}: A display of the Spearman's (bottom right half) and Pearson's (top left part) correlation matrix between Facebook metrics and information from the Cartesio experience. The colour of the hexagons indicates the correlation value, blue for positive and red for negative correlations. The size of the hexagon indicates the p-value of the correlation.

\section*{Table Legends}

\textbf{Table \ref{tab:demo}}: Descriptive statistics table showing the demographic characteristics of participants. Data marked (\textbf{*}) includes post-university qualifications, as separate data was not available for Italian demographics. The ``Frequency'' column presents the actual count of participants within each category, while the ``Known $\%$'' column reflects the proportion of these participants among those who provided their demographic data. The ``Italian Demographic $\%$'' column offers a baseline for comparison, sourced from national statistics.


\begin{thebibliography}{10}
\bibliographystyle{plain}
\bibitem{audiweb2020}
Audiweb database.
\newblock \url{https://www.audiweb.it/dati/index.html?anno=2020}, 2020.

\bibitem{baldassarri2008partisans}
Delia Baldassarri and Andrew Gelman.
\newblock Partisans without constraint: Political polarization and trends in american public opinion.
\newblock {\em American Journal of Sociology}, 114(2):408--446, 2008.

\bibitem{baumann2021emergence}
Fabian Baumann, Philipp Lorenz-Spreen, Igor~M Sokolov, and Michele Starnini.
\newblock Emergence of polarized ideological opinions in multidimensional topic spaces.
\newblock {\em Physical Review X}, 11(1):011012, 2021.

\bibitem{brenan2022americans}
M~Brenan.
\newblock Americans’ trust in media remains near record low, 2022.

\bibitem{brugnoli2019recursive}
Emanuele Brugnoli, Matteo Cinelli, Walter Quattrociocchi, and Antonio Scala.
\newblock Recursive patterns in online echo chambers.
\newblock {\em Scientific Reports}, 9(1):20118, 2019.

\bibitem{chenpublic}
Yijing Chen, Zolt{\'a}n Kmetty, Gerardo I{\~n}iguez, and Elisa Omodei.
\newblock The public that engages invisibly: What visible engagement fails to capture in online political communication.

\bibitem{cinelli2020selective}
Matteo Cinelli, Emanuele Brugnoli, Ana~Lucia Schmidt, Fabiana Zollo, Walter Quattrociocchi, and Antonio Scala.
\newblock Selective exposure shapes the facebook news diet.
\newblock {\em PloS one}, 15(3):e0229129, 2020.

\bibitem{cinelli2021dynamics}
Matteo Cinelli, Andra{\v{z}} Pelicon, Igor Mozeti{\v{c}}, Walter Quattrociocchi, Petra~Kralj Novak, and Fabiana Zollo.
\newblock Dynamics of online hate and misinformation.
\newblock {\em Scientific reports}, 11(1):22083, 2021.

\bibitem{Const2121cane}
Italian~Constitutional Court.
\newblock Italian constitutional court, sentenza n.150.
\newblock \url{https://www.giurisprudenzapenale.com/wp-content/uploads/2021/07/pronuncia_150_2021.pdf}, july 12 2021.

\bibitem{dimaggio1996have}
Paul DiMaggio, John Evans, and Bethany Bryson.
\newblock Have american's social attitudes become more polarized?
\newblock {\em American journal of Sociology}, 102(3):690--755, 1996.

\bibitem{ecker2022psychological}
Ullrich~KH Ecker, Stephan Lewandowsky, John Cook, Philipp Schmid, Lisa~K Fazio, Nadia Brashier, Panayiota Kendeou, Emily~K Vraga, and Michelle~A Amazeen.
\newblock The psychological drivers of misinformation belief and its resistance to correction.
\newblock {\em Nature Reviews Psychology}, 1(1):13--29, 2022.

\bibitem{eveland2003impact}
William~P Eveland~Jr and Dhavan~V Shah.
\newblock The impact of individual and interpersonal factors on perceived news media bias.
\newblock {\em Political psychology}, 24(1):101--117, 2003.

\bibitem{fico2004influence}
Frederick Fico, John~D Richardson, and Steven~M Edwards.
\newblock Influence of story structure on perceived story bias and news organization credibility.
\newblock {\em Mass communication \& society}, 7(3):301--318, 2004.

\bibitem{WEF2024}
World~Economic Forum.
\newblock Global risks report 2024.
\newblock Technical report, World Economic Forum, 2024.

\bibitem{freeman2020measuring}
Cole Freeman, Hamed Alhoori, and Murtuza Shahzad.
\newblock Measuring the diversity of facebook reactions to research.
\newblock {\em Proceedings of the ACM on Human-Computer Interaction}, 4(GROUP):1--17, 2020.

\bibitem{Cartesio}
Pietro Gravino.
\newblock {S}ony {CSL}'s {C}artesio {P}latform.
\newblock \url{https://cartesio.news}, 2020.

\bibitem{gutmann2009democracy}
Amy Gutmann and Dennis~F Thompson.
\newblock {\em Democracy and disagreement}.
\newblock Harvard University Press, 2009.

\bibitem{jones2015declining}
David~R Jones.
\newblock Declining trust in congress: Effects of polarization and consequences for democracy.
\newblock In {\em The Forum}, volume~13, pages 375--394. De Gruyter, 2015.

\bibitem{kattenbeck2019understanding}
Markus Kattenbeck and David Elsweiler.
\newblock Understanding credibility judgements for web search snippets.
\newblock {\em Aslib Journal of Information Management}, 71(3):368--391, 2019.

\bibitem{kidd2023}
Celeste Kidd and Abeba Birhane.
\newblock How ai can distort human beliefs.
\newblock {\em Science}, 380(6651):1222--1223, 2023.

\bibitem{koudenburg2021new}
Namkje Koudenburg, Henk~AL Kiers, and Yoshihisa Kashima.
\newblock A new opinion polarization index developed by integrating expert judgments.
\newblock {\em Frontiers in psychology}, 12:738258, 2021.

\bibitem{layman2006party}
Geoffrey~C Layman, Thomas~M Carsey, and Juliana~Menasce Horowitz.
\newblock Party polarization in american politics: Characteristics, causes, and consequences.
\newblock {\em Annu. Rev. Polit. Sci.}, 9:83--110, 2006.

\bibitem{lee2022social}
Amber Hye-Yon Lee.
\newblock Social trust in polarized times: how perceptions of political polarization affect americans’ trust in each other.
\newblock {\em Political behavior}, 44(3):1533--1554, 2022.

\bibitem{lewandowsky2024}
Stephan Lewandowsky, Ullrich K.~H. Ecker, John Cook, Sander van~der Linden, Jon Roozenbeek, Naomi Oreskes, and Lee~C. McIntyre.
\newblock Liars know they are lying: differentiating disinformation from disagreement.
\newblock {\em Humanities and Social Sciences Communications}, 11(1):986, 2024.

\bibitem{michael2021relationship}
Robert~B Michael and Brooke~O Breaux.
\newblock The relationship between political affiliation and beliefs about sources of “fake news”.
\newblock {\em Cognitive research: principles and implications}, 6:1--15, 2021.

\bibitem{muhammed2022disaster}
Sadiq Muhammed~T and Saji~K Mathew.
\newblock The disaster of misinformation: a review of research in social media.
\newblock {\em International journal of data science and analytics}, 13(4):271--285, 2022.

\bibitem{Reuters2023}
Newman N., Fletcher R., Eddy K.C.T., and Nielsen R.K.
\newblock Digital news report 2023.
\newblock Technical report, Oxford: Reuters Institute for the Study of Journalism, 2023.

\bibitem{Reuters2019}
Newman N., Fletcher R., and Nielsen R.K.
\newblock Reuters institute digital news report 2019.
\newblock Technical report, Oxford: Reuters Institute for the Study of Journalism, 2023.

\bibitem{osmundsen2021partisan}
Mathias Osmundsen, Alexander Bor, Peter~Bjerregaard Vahlstrup, Anja Bechmann, and Michael~Bang Petersen.
\newblock Partisan polarization is the primary psychological motivation behind political fake news sharing on twitter.
\newblock {\em American Political Science Review}, 115(3):999--1015, 2021.

\bibitem{peach2024seeing}
Kaitlin Peach, Joseph Ripberger, Kuhika Gupta, Andrew Fox, Hank Jenkins-Smith, and Carol Silva.
\newblock Seeing lies and laying blame: Partisanship and us public perceptions about disinformation.
\newblock {\em Harvard Kennedy School Misinformation Review}, 2024.

\bibitem{pennycook2019lazy}
Gordon Pennycook and David~G Rand.
\newblock Lazy, not biased: Susceptibility to partisan fake news is better explained by lack of reasoning than by motivated reasoning.
\newblock {\em Cognition}, 188:39--50, 2019.

\bibitem{popoviciu1935equations}
Tiberiu Popoviciu.
\newblock Sur les {\'e}quations alg{\'e}briques ayant toutes leurs racines r{\'e}elles.
\newblock {\em Mathematica}, 9(129-145):20, 1935.

\bibitem{powers2019shouting}
Elia Powers, Michael Koliska, and Pallavi Guha.
\newblock “shouting matches and echo chambers”: Perceived identity threats and political self-censorship on social media.
\newblock {\em International Journal of Communication}, 13:20, 2019.

\bibitem{prochaska2023mobilizing}
Stephen Prochaska, Kayla Duskin, Zarine Kharazian, Carly Minow, Stephanie Blucker, Sylvie Venuto, Jevin~D West, and Kate Starbird.
\newblock Mobilizing manufactured reality: How participatory disinformation shaped deep stories to catalyze action during the 2020 us presidential election.
\newblock {\em Proceedings of the ACM on Human-Computer Interaction}, 7(CSCW1):1--39, 2023.

\bibitem{rapp2016moral}
Carolin Rapp.
\newblock Moral opinion polarization and the erosion of trust.
\newblock {\em Social science research}, 58:34--45, 2016.

\bibitem{reinero2023partisans}
Diego~A Reinero, Elizabeth~Ann Harris, Steve Rathje, Annie Duke, and Jay~J Van~Bavel.
\newblock Partisans are more likely to entrench their beliefs in misinformation when political outgroup members fact-check claims, 2023.

\bibitem{roy2022systematic}
Deepjyoti Roy and Mala Dutta.
\newblock A systematic review and research perspective on recommender systems.
\newblock {\em Journal of Big Data}, 9(1):59, 2022.

\bibitem{schultz1998reviving}
Julianne Schultz.
\newblock {\em Reviving the fourth estate: Democracy, accountability and the media}.
\newblock Cambridge University Press, 1998.

\bibitem{spitale2023}
Giovanni Spitale, Nikola Biller-Andorno, and Federico Germani.
\newblock Ai model gpt-3 (dis)informs us better than humans.
\newblock {\em Science Advances}, 9(26):eadh1850, 2023.

\bibitem{tokita2021polarized}
Christopher~K Tokita, Andrew~M Guess, and Corina~E Tarnita.
\newblock Polarized information ecosystems can reorganize social networks via information cascades.
\newblock {\em Proceedings of the National Academy of Sciences}, 118(50):e2102147118, 2021.

\bibitem{wang2019systematic}
Yuxi Wang, Martin McKee, Aleksandra Torbica, and David Stuckler.
\newblock Systematic literature review on the spread of health-related misinformation on social media.
\newblock {\em Social science \& medicine}, 240:112552, 2019.

\end{thebibliography}
\end{document}


\maketitle

\section*{Introduction}
This Supporting Information complements the findings of ``From trust in news to disagreement: is misinformation more controversial?'' It provides additional analyses, and supplementary figures and tables to enhance the transparency, reproducibility, and understanding of the research. Refer to this document while reviewing the main text.

\section*{Outline of Supporting Information}

\begin{enumerate}
  \item \textbf{Summary statistics of Cartesio news articles}
  \item \textbf{Disagreement metric comparison}
  \item \textbf{Trust determinants}
  \item \textbf{Facebook Metrics Analysis}
  \item \textbf{References}
\end{enumerate}

\section{Summary statistics of Cartesio news articles}
\begin{figure}[ht]
    \centering
    \includegraphics[width=.8\textwidth]{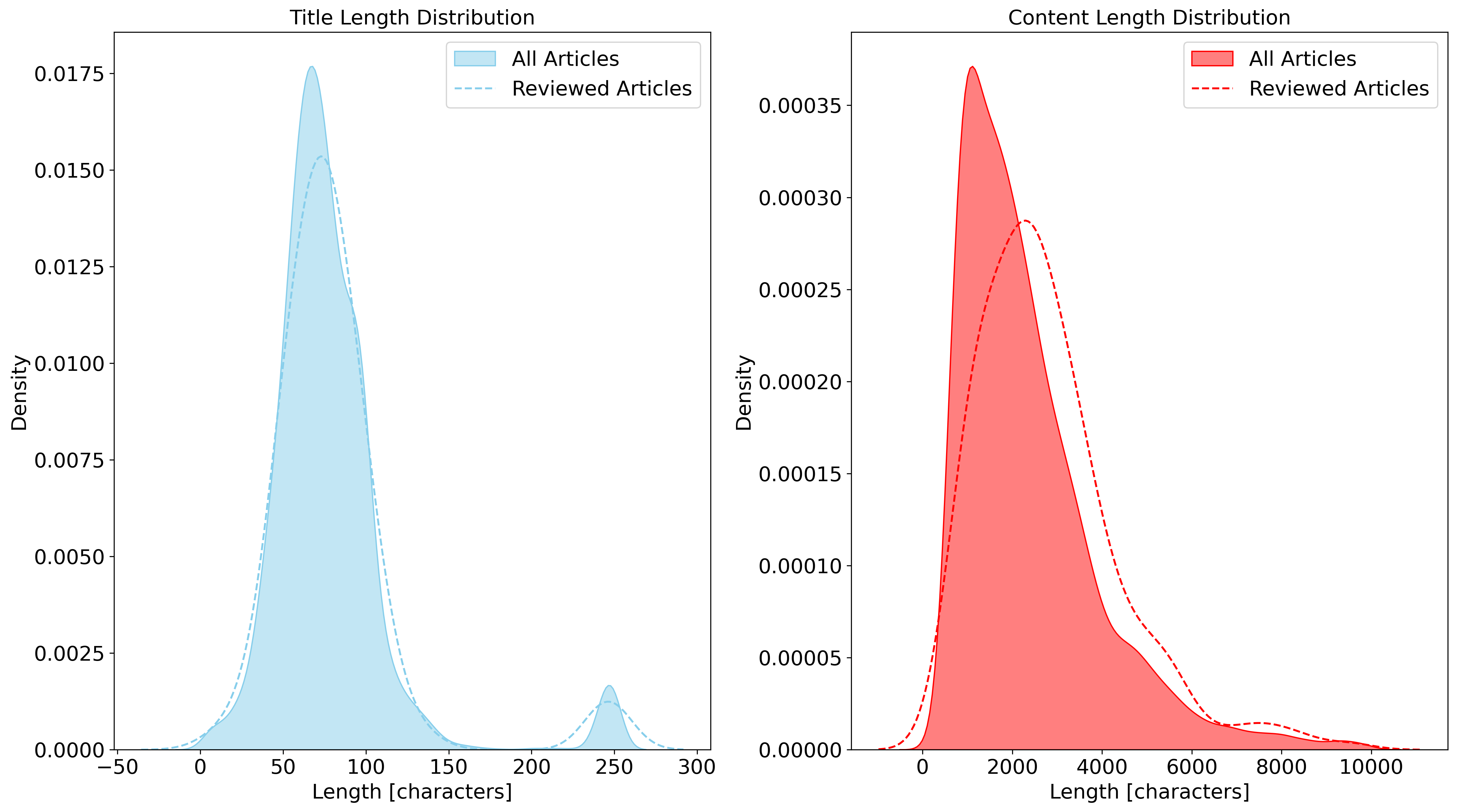}
    \caption{The panels display the length in characters of news articles in the Cartesio experiment data. \textbf{Left:} The panel shows the length distribution of titles. The shaded area displays the distribution of title lengths in the whole dataset, while the dotted line shows the title lengths of articles reviewed by participants in the experiment. \textbf{Right:} the panel shows the distribution of content length of articles in the Cartesio experiment data, the shaded area presents the distribution over all articles in the dataset, while the dotted line shows the length of the articles reviewed by participants.}
    \label{fig:Len-Dist}
\end{figure}

In this section we provide an overview of the characteristics of news articles present in the Cartesio dataset. When not explicitly stated, statistics refer to news articles that where reviewed at least 4 times by participants, since articles that received less reviews did not contribute to the analysis presented in the paper.
In Figure~\ref{fig:Len-Dist}, we show the length distribution of title and content fields of the articles. The majority of titles falls between 50 and 100 characters in length (with a mean of 78 characters of length), with a secondary peak around 250 characters, consistent with most titles being comprised of one sentence and a small minority being comprised of two sentences.
Content lengths have a mean of 2360 characters. Reviewed articles have a slightly higher mean with and average length of 2780 characters.

News categories where provided by AGCOM and imputed using a topic modelling procedure. The category names in English translate to:
\begin{itemize}
  \item \textbf{Ambiente} - Environment
  \item \textbf{Diritti della persona} - Human Rights
  \item \textbf{Economia e lavoro} - Economy and Work
  \item \textbf{Immigrazione} - Immigration
  \item \textbf{Infrastrutture e mobilità} - Infrastructure and Mobility
  \item \textbf{Istituzioni e politica} - Institutions and Politics
  \item \textbf{Istruzione} - Education
  \item \textbf{Salute} - Health
  \item \textbf{Scienza e tecnologia} - Science and Technology
  \item \textbf{Sicurezza e giustizia} - Security and Justice
\end{itemize}

\begin{figure}[ht]
    \centering
    \includegraphics[width=.8\textwidth]{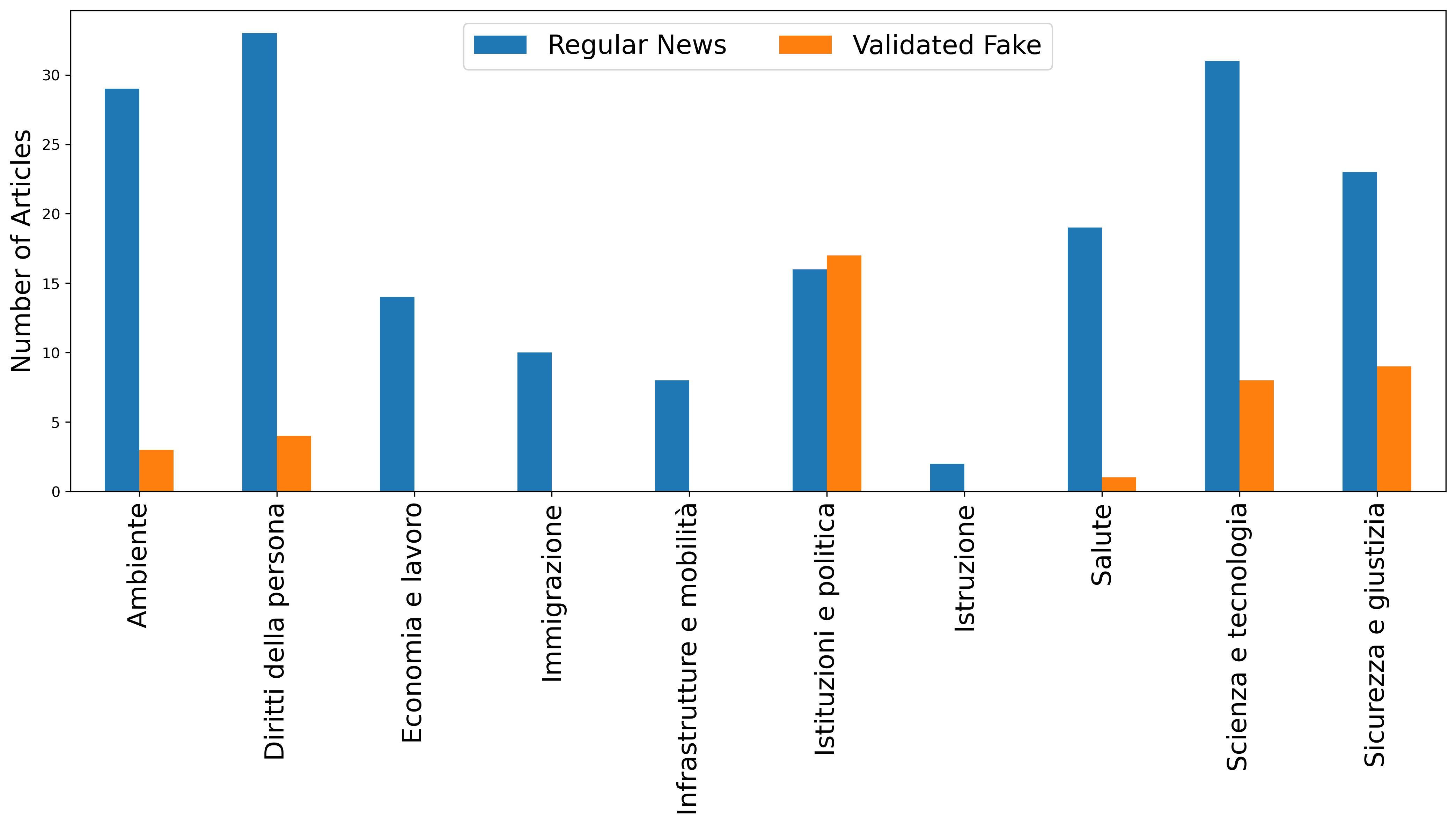}
    \caption{Distribution of articles across categories as imputed by a topic modelling procedure. The blue bars display the number of articles that where not labelled as fake by expert evaluators, whole the orange bars show the number of articles that where labelled as fake.}
    \label{fig:Cat-Dist}
\end{figure}

The distribution of articles in the imputed categories can be found in Figure~\ref{fig:Cat-Dist}, with news labelled as \emph{fake} being separated from the rest in the orange bars.

\begin{figure}
    \centering
    \includegraphics[width=\textwidth]{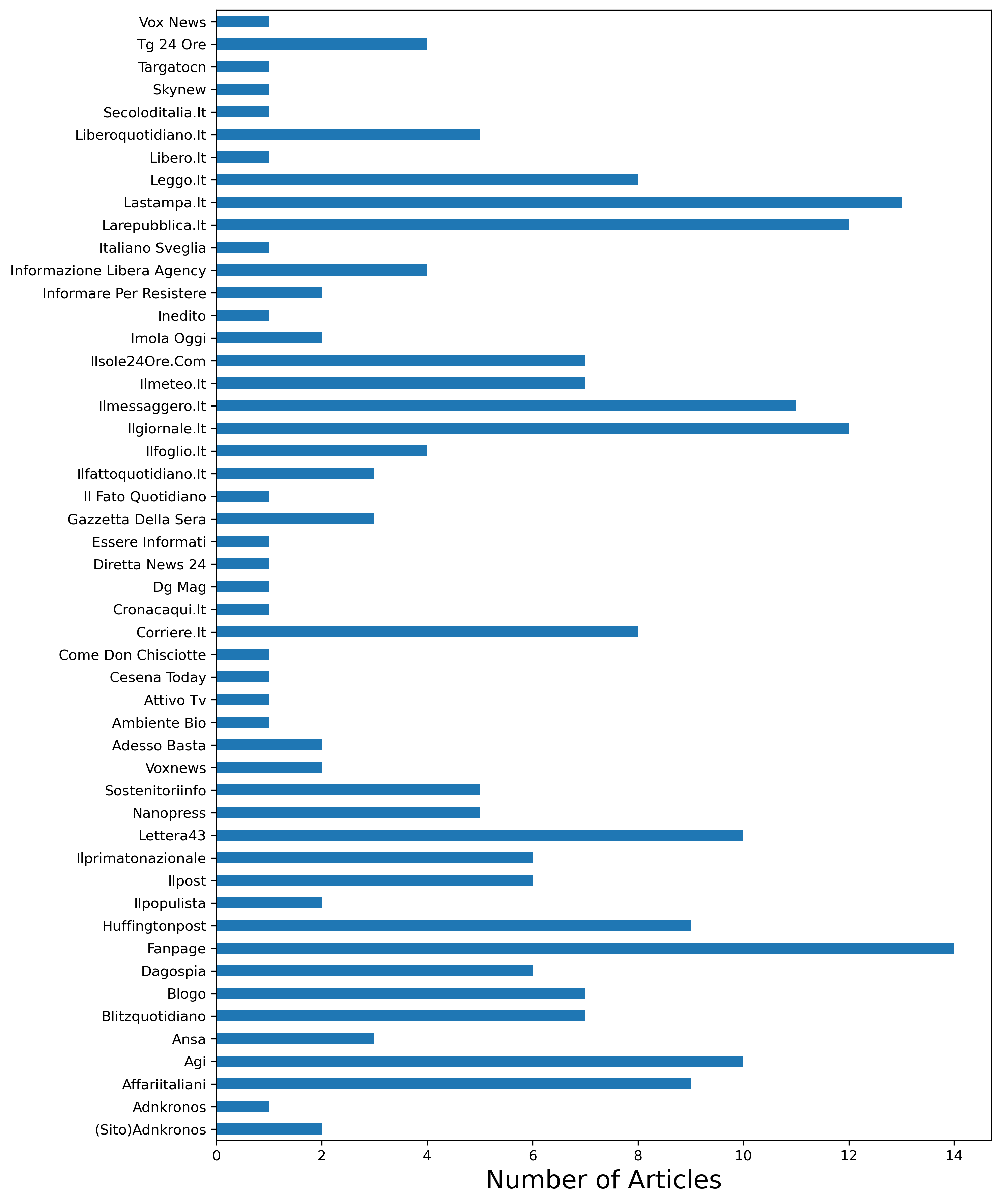}
    \caption{The number of articles by news sources that were reviewed by participants during the experiment.}
    \label{fig:Sorc-Dist}
\end{figure}

\begin{figure}
    \centering
    \includegraphics[width=0.5\textwidth]{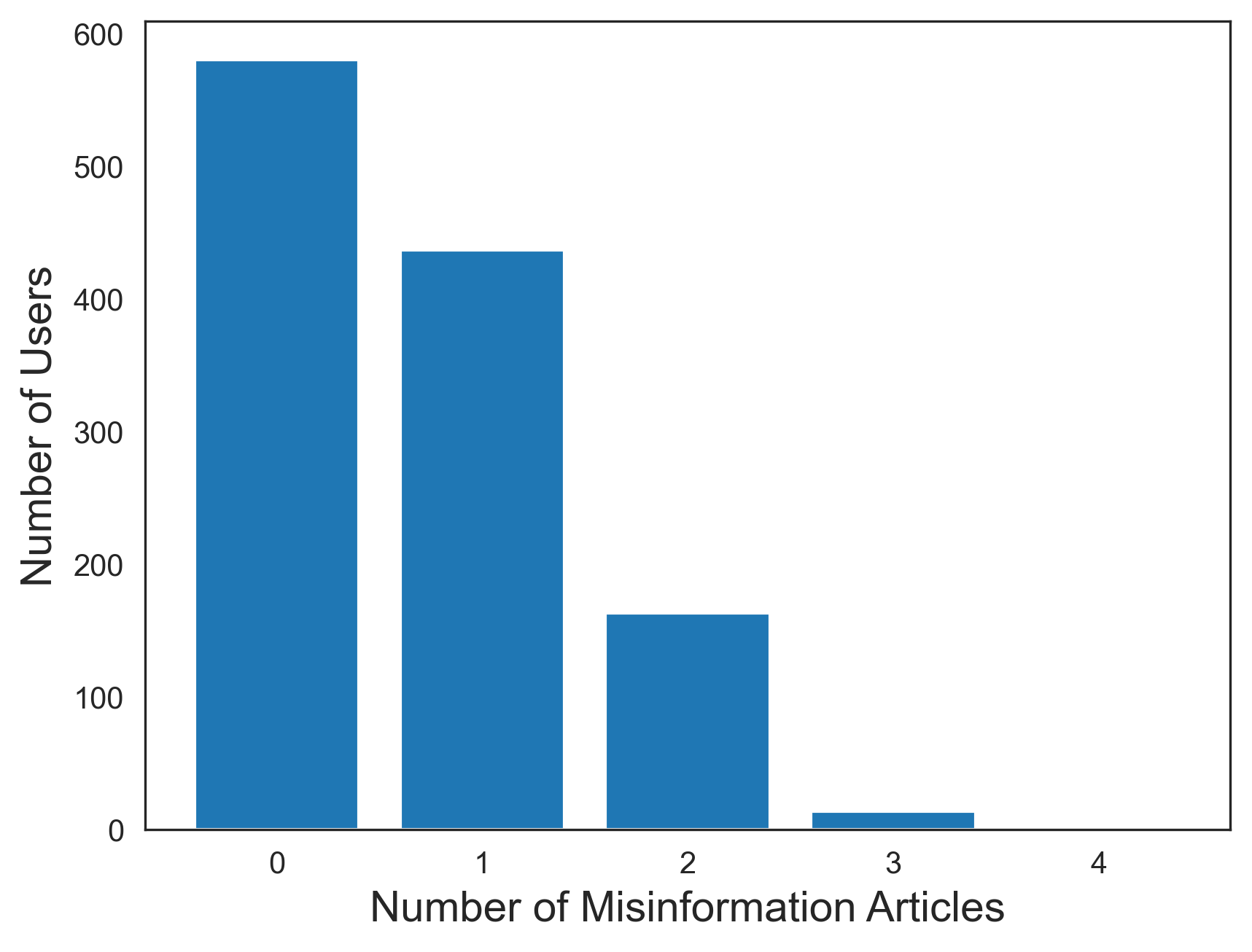}
    \caption{The frequency with which news items labelled as misinformation were among the rated items for each user.}
    \label{fig:misinfodist}
\end{figure}

Figure~\ref{fig:Sorc-Dist} displays the distribution of sources in news articles reviewed by participants.
Finally, Figure~\ref{fig:misinfodist} displays the the frequency with which users rated one, two, or more news items labelled as `validated fake', by the expert evaluators.

\section{Disagreement metric comparison}
In this section we provide an overview of the different metrics investigated the distributional properties of trust.
Table~\ref{tab:trust statistics} displays the main characteristics of the trust distribution, divided into `Misinformation' and regular news items, as illustrated in Figure 2 of the manuscript.
\begin{table}[ht]
    \centering
    \footnotesize
\begin{tabular}{l|rrrrrrrr}
\toprule
{} & \multicolumn{8}{l}{Trust} \\
{} &    count &  mean &   std &  min &   25\% &   50\% &   75\% &    max \\
Misinformation &          &       &       &      &       &       &       &        \\
\midrule
No             & 16018.00 & 61.60 & 29.25 & 0.00 & 40.00 & 70.00 & 85.00 & 100.00 \\
Yes            &  1157.00 & 36.07 & 33.12 & 0.00 &  5.00 & 25.00 & 65.00 & 100.00 \\
\bottomrule
\end{tabular}
    \caption{The main characteristics of the trust distribution, divided into `Misinformation' and regular news items.}
    \label{tab:trust statistics}
\end{table}

\begin{figure}
    \centering
    \includegraphics[width=\textwidth]{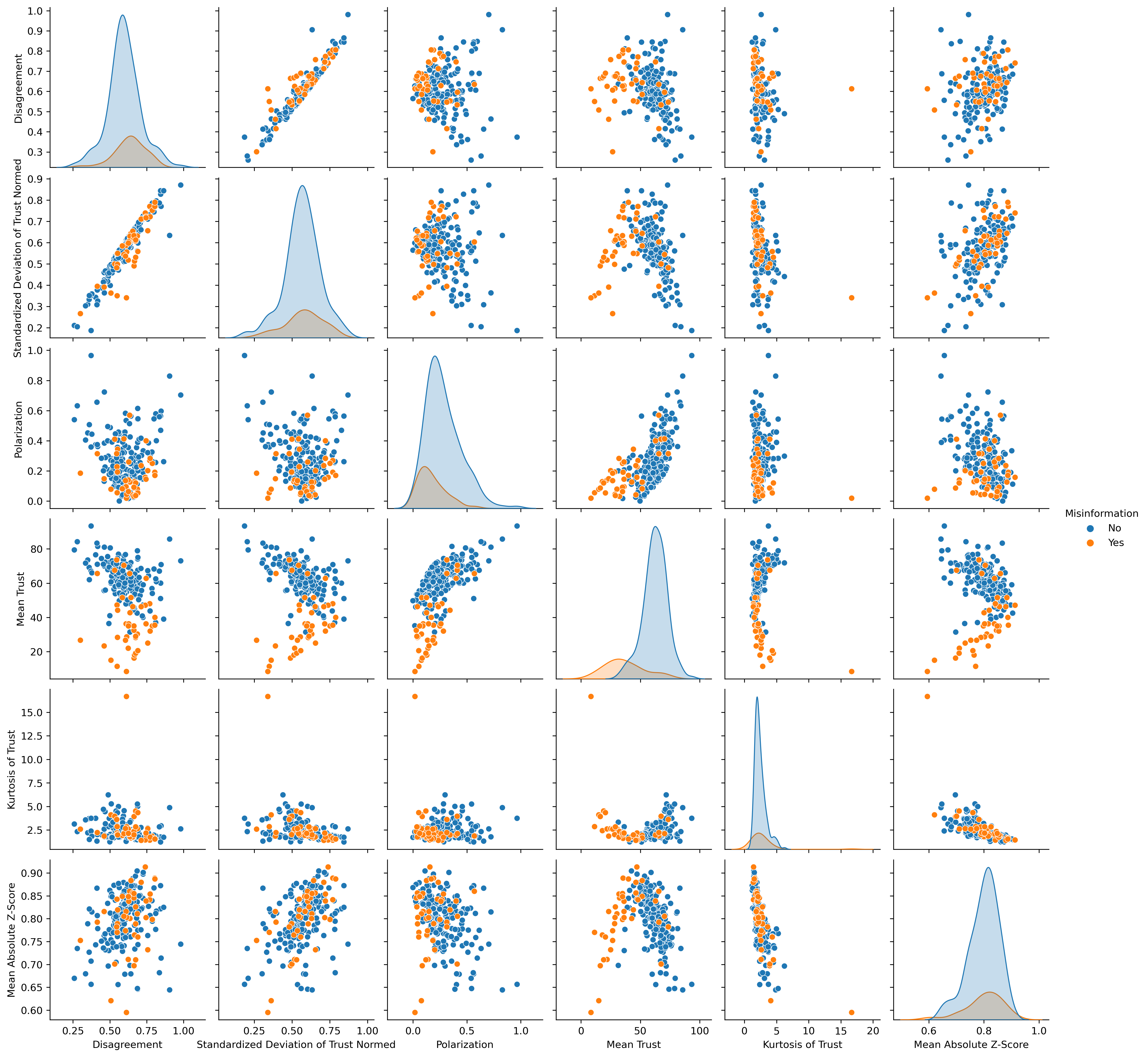}
    \caption{The pair-plot encapsulates pairwise scatter plots for quantitative comparison, with kernel density estimations on the diagonal showing the distribution of individual metrics. Each point represents an observation, colour-coded as blue for `No Misinformation' and orange for `Yes Misinformation'.}
    \label{fig:metrics}
\end{figure}

Beyond the \emph{disagreement} metric described in the paper (Section Disagreement and Misinformation) we investigated a list of possible alternatives, namely: the standard deviation of trust scores per news item (Standard Deviation of Trust Normed), distributional polarisation as formulated in \cite{duclos2006polarization} (Polarization), the kurtosis index and the mean absolute value of the Z-score of news items (Mean Absolute Z-score).
Figure~\ref{fig:metrics} shows a pairwise comparison between the various metrics, along with kernel density estimations for each metric, highlighting the distribution of data for each variable considered.

\begin{figure}
    \centering
    \includegraphics[width=\textwidth]{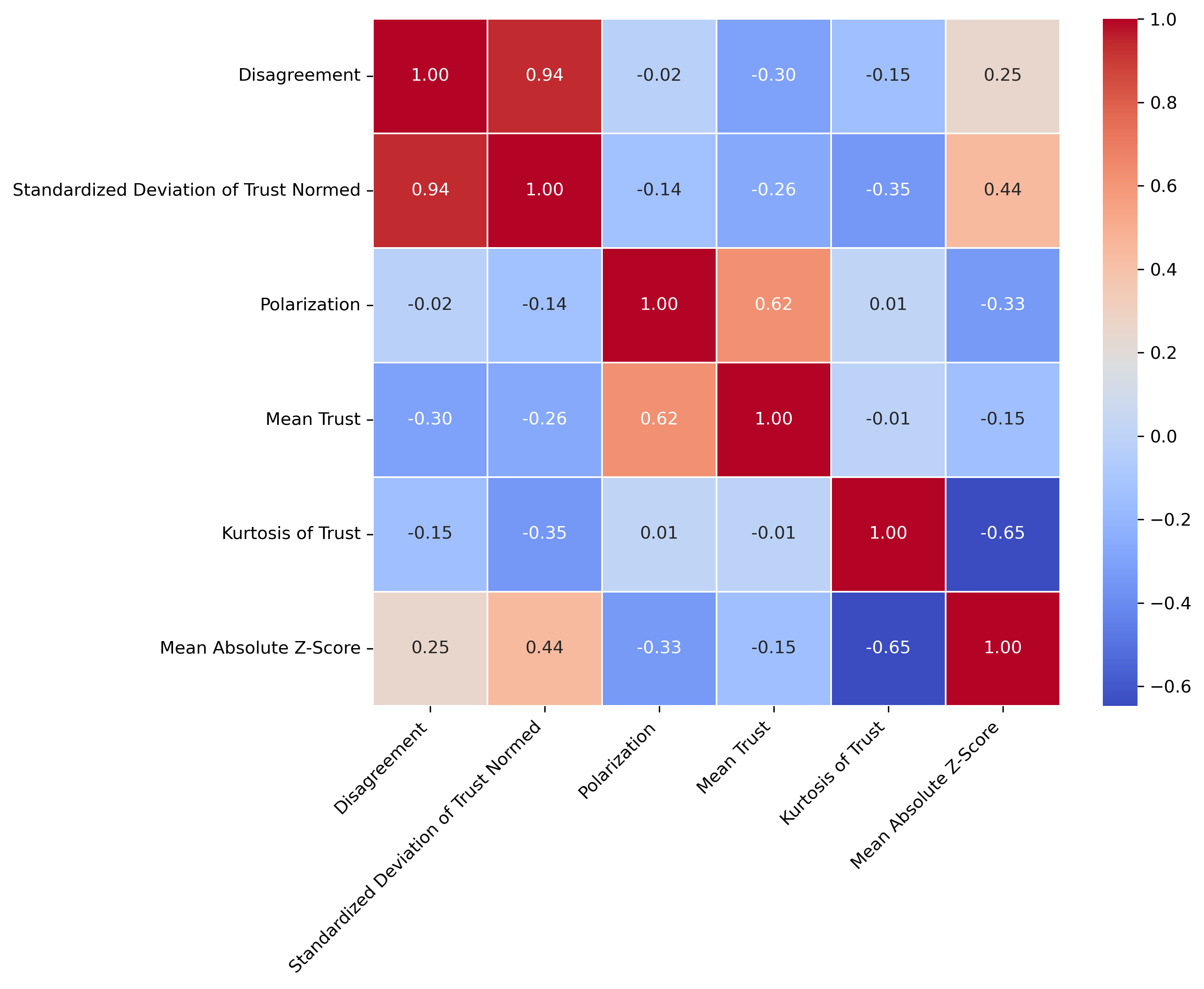}
    \caption{Pearson correlation values for the metrics proposed to measure disagreement and their relation to `Mean Trust'.}
    \label{fig:corrplot}
\end{figure}

Figure~\ref{fig:corrplot} shows how \emph{disagreement} and Standard deviation closely correlate, with \emph{disagreement} correcting for the parabolic dependency on the mean. Further, `Mean Absolute Z-score' correlates with disagreement and standard deviation, as is to be expected. This last measure is commonly used in measures of polarisation.
The measure identified as `Polarisation' is a strong candidate for our needs, and is defined in such a way as to follow specific properties (as argued in \cite{duclos2006polarization}) that correspond to the expected behaviour of polarisation, but the lack of correlation with standardised deviation suggests that in a context with data scarcity it doesn't behave according to the distributional characteristic expected by Di Maggio, et. al. \cite{dimaggio1996have}.
Finally, Kurtosis could signify stronger tails, that could be expected in bimodal distributions, but again, the lack of correlation with standard deviation poses difficulties in interpretation that might come from the lack of normality in the observed distributions.

\section{Trust determinants}
Since in principle trust scores could be influenced by the sequence of articles reviewed by readers we investigate the influence of reading time and relative position to first news item labelled ad fake on trust ratings.

\begin{figure}[ht]
    \centering
    \includegraphics[width=.8\textwidth]{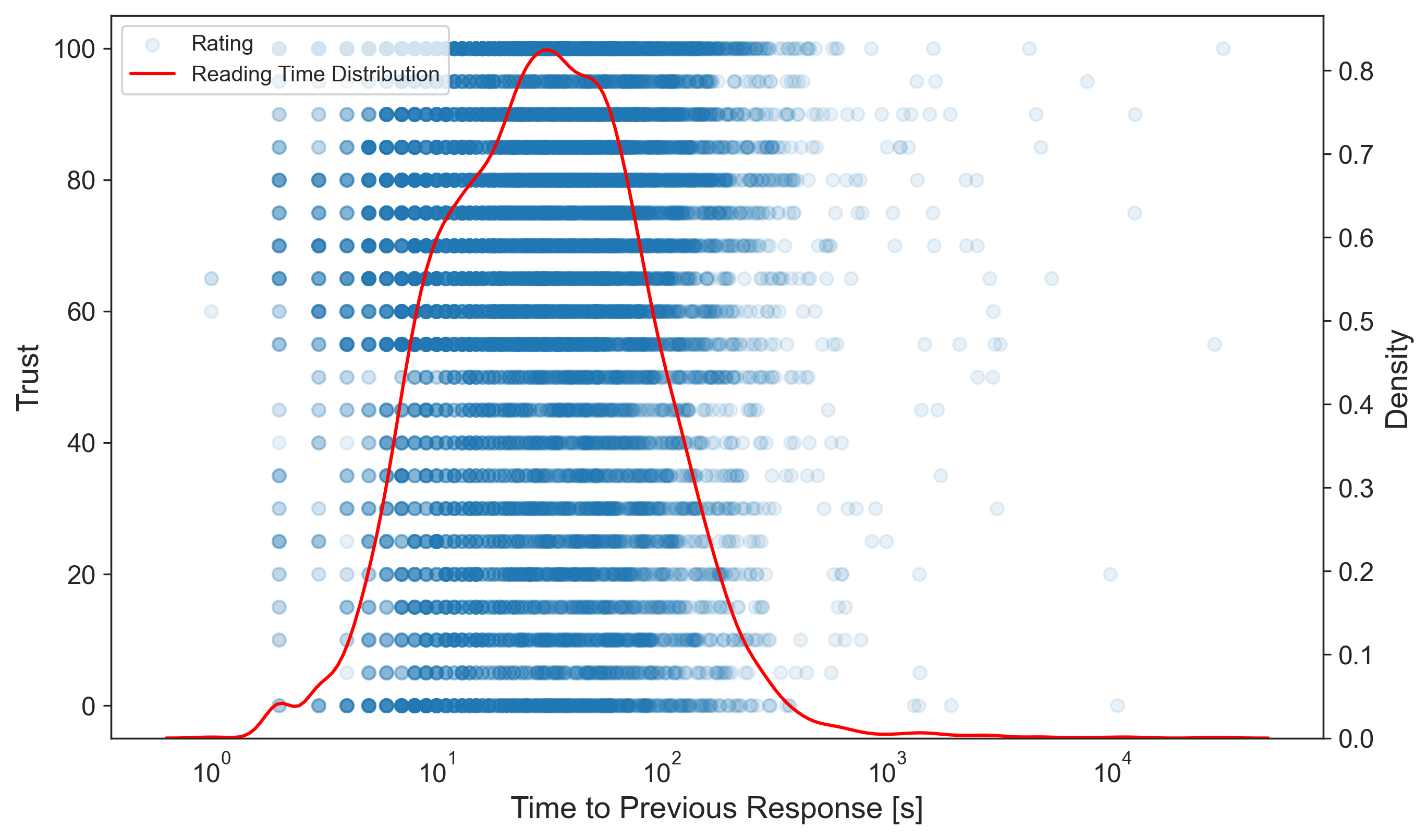}
    \caption{The scatter-plot displays the rating of users. On the x-axis the time from the previous rating in the same session is used as a proxy for reading time. On the left y-axis the trust scores received by the news item. The red line indicates the distribution of rating times and is associated with the right y-axis.}
    \label{fig:times}
\end{figure}

In Figure \ref{fig:times} we display the trust ratings as a function of time between ratings (as a proxy of reading time). 
Additionally, the figure displays the distribution of rating times. We observe a mean reading time of 63 seconds and no significant dependency between the two variables.

The Pearson correlation analysis between reading time and trust ratings yielded a coefficient of 0.013, indicating a negligible linear relationship. Additionally, the p-value of 0.109 suggests that this weak correlation is not statistically significant, reinforcing the conclusion that reading time does not reliably predict trust ratings.

\begin{figure}[h]
    \centering
    \includegraphics[width=0.5\textwidth]{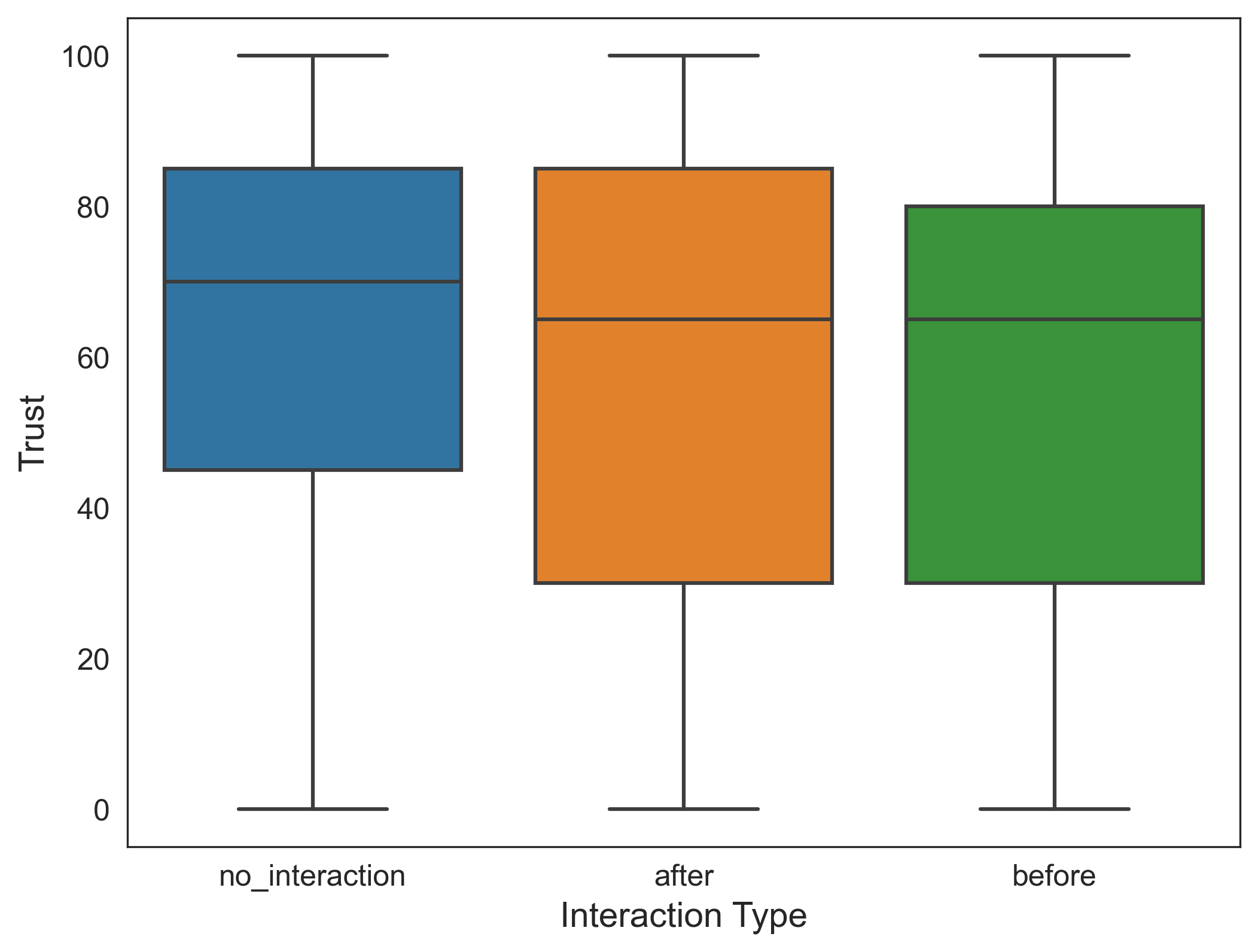}
    \caption{The box-plots display the distribution of trust ratings before, after and for participants who did not interact with news items labelled as `validated fake'.}
    \label{fig:type}
\end{figure}

Finally, in Figure~\ref{fig:type} we display the trust score distribution of news items reviewed before the first interaction of a user with an item labelled as fake.
To test the probability with which we can reject the hypothesis that interaction with news items labelled as `validated fake' has no effect on participants we compute the Mann-Whitney U test between the distributions `before' and `after' (visible in Figure~\ref{fig:type}) and we obtain a p-value of 0.26, which indicates we cannot statistically reject the hypothesis.

\begin{figure}[ht]
    \centering
    \includegraphics[width=.8\textwidth]{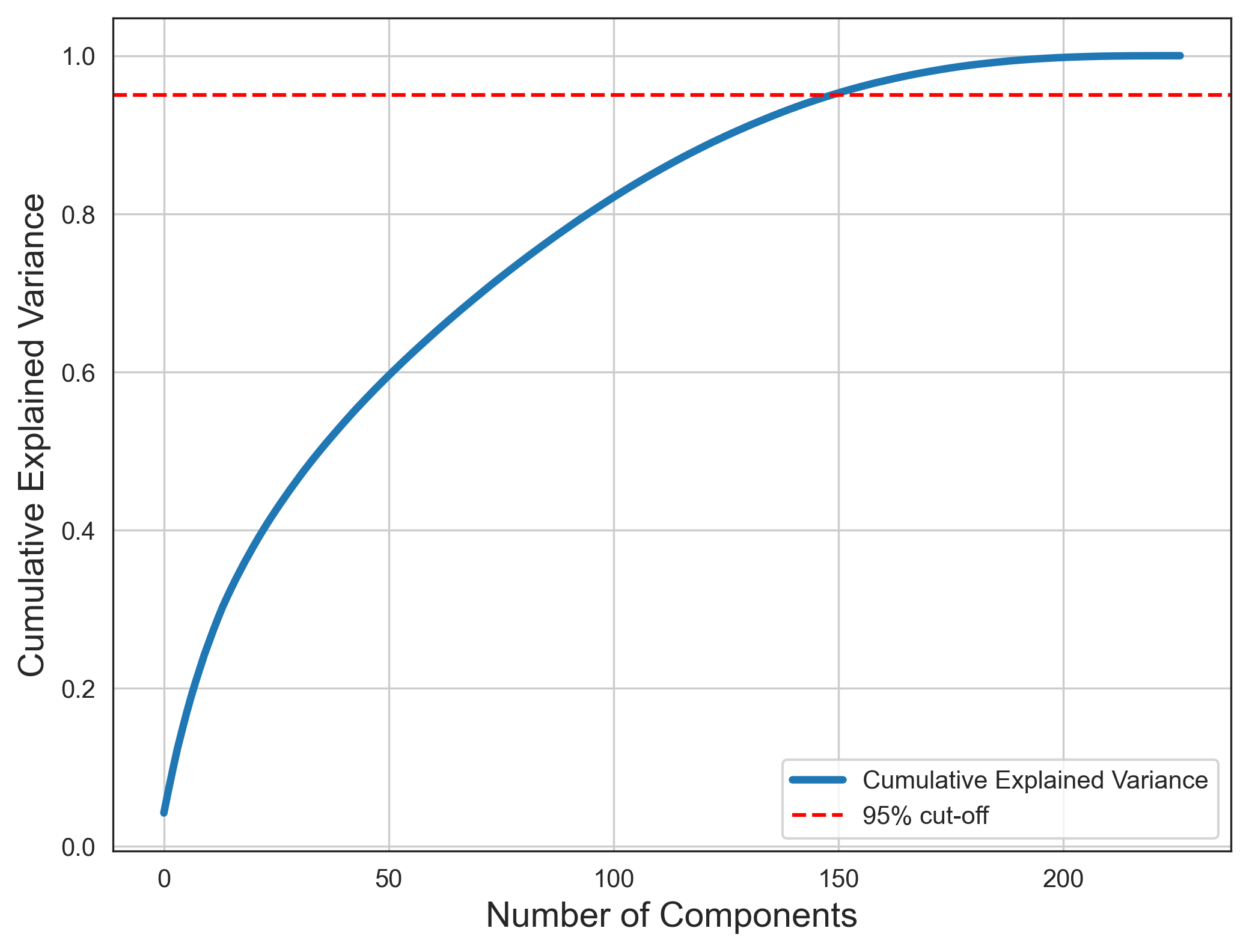}
    \caption{The cumulative explained variance as a result of a principal component analysis (PCA), as a function of the number of components. The red dotted line indicates the level of $95\%$ explained variance.}
    \label{fig:pca}
\end{figure}

In order to check for determinants to the structure of trust ratings we perform a principal component analysis (PCA) on the scores given by users. The rating matrix was filled using the mean value for each article as an imputed value for the rating of that user on that article. Further, ratings were standardised before performing the PCA. As shown in Figure~\ref{fig:pca}, the cumulative explained variance as a function of the number of components follows a very regular pattern, without elbows. Additionally, to explain $95\%$ of the variance, we would need 150 components, indicating a high level of diversity among users that cannot be explained with a small number of groups.

\begin{figure}[ht]
    \centering
    \includegraphics[width=0.5\linewidth]{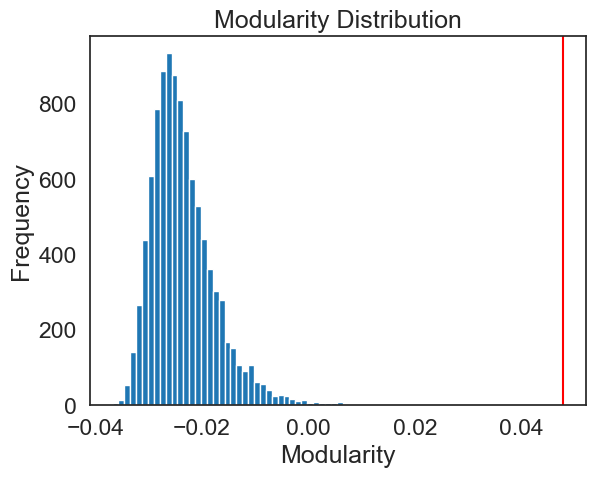}
    \caption{The distribution of modularity scores of reshuffled community labels in blue, and the modularity score of the optimized Louvain algorithm in red.}
    \label{fig:louv}
\end{figure}

We also look for structure in the bipartite network of ratings. To do this we constructed a network of positive ratings, filtering on positive values of the rating z-score. We then perform an optimised Louvain community detection. The algorithm finds two distinct communities. We perform a bootstrap analysis comparing the modularity of the communities to a random reshuffling of labels and find the modularity to be significantly higher than the random case, although not very high in absolute terms (see Figure~\ref{fig:louv}). The two structural analyses taken together suggest a diverse set of users, who's rating are not easily reducible to partisan structures, that however recognises partisanship in the sources and is influenced by this perception.

\begin{figure}
    \centering
    \includegraphics[width=.9\textwidth]{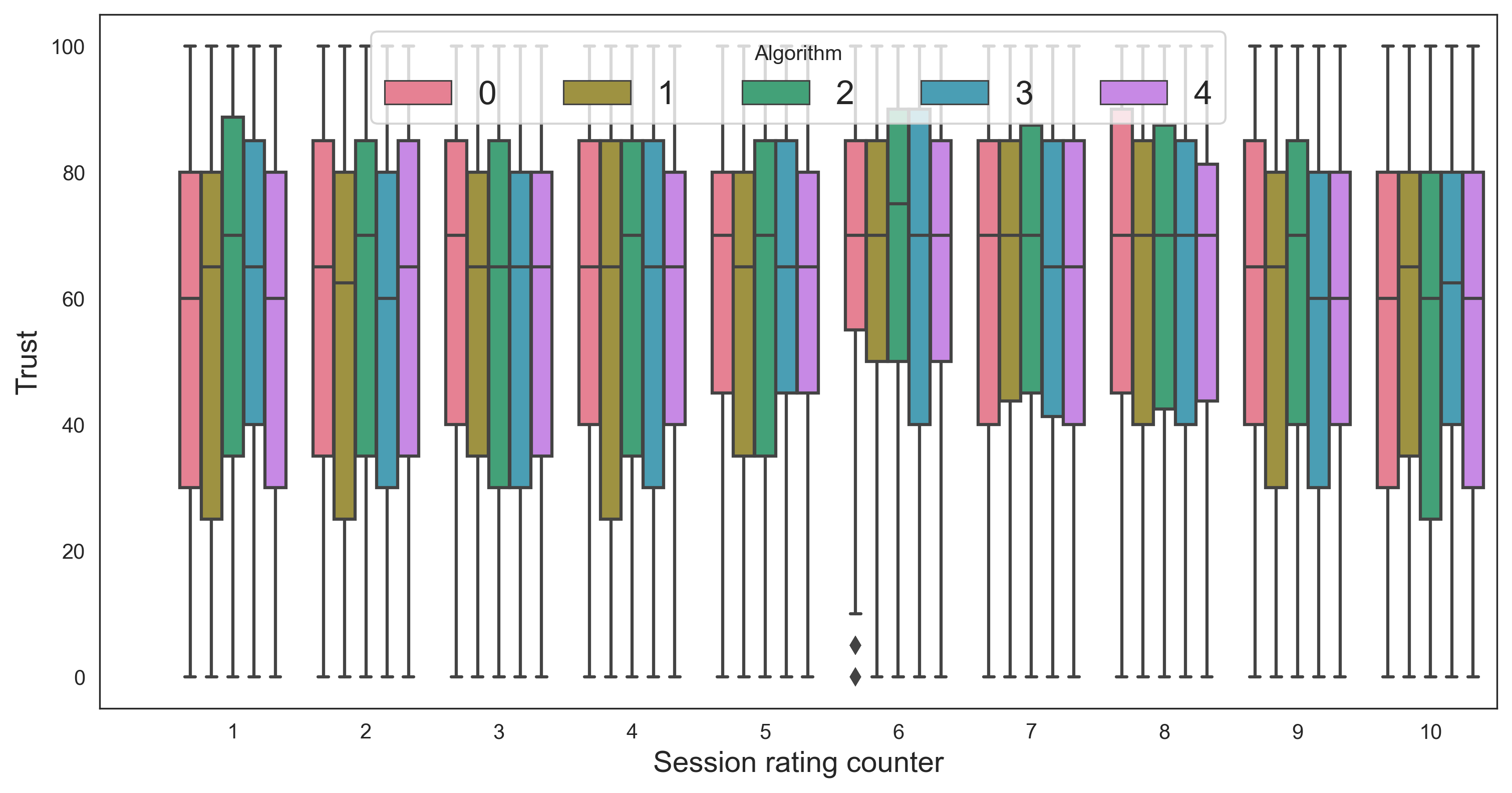}
    \caption{Distribution of trust ratings as a function of the number of ratings given up to that point by the user (\emph{Session rating counter}). Different recommender-algorithms are displayed as differently coloured box-plots. The boxes represent the area between the 25th and 75th percentile. The lines within the box are the median value, while the outer bars display the area between the 5th and 95th percentile.}
    \label{fig:algo}
\end{figure}

Since users are assigned recommender algorithms at random by the Cartesio platform, we check whether the recommender influences the rating distributions. In Figure~\ref{fig:algo} we display the rating distribution as a function of the position in the session sequence of a rating. Different algorithms are displayed as differently coloured box-plots. A visual inspection uncovers no systematic effect of the algorithm in the first 10 ratings. Beyond the tenth rating data becomes scarce and is no longer easily comparable.
\begin{figure}
    \centering
    \includegraphics[width=.8\textwidth]{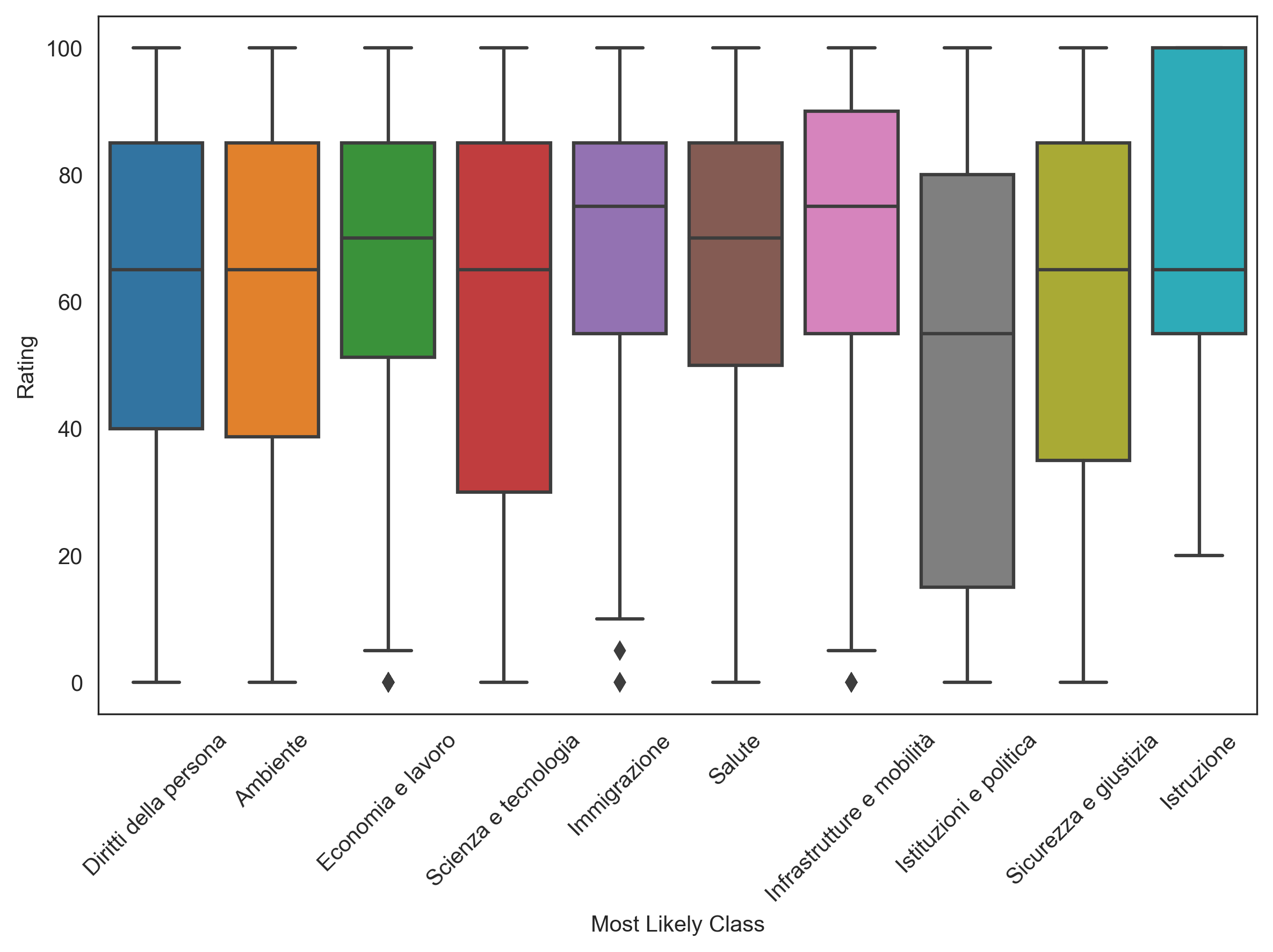}
    \caption{Boxplot showing the distribution of ratings for items assigned the most likely topic.}
    \label{fig:topic-trust}
\end{figure}

In order to check for effects due to the news item topic we report the trust distributions associated with the most likely topic for each article included in the analysis (see Figure~\ref{fig:topic-trust} for a summary). We get significant effects on the topics `Economia e lavoro', `Scienza e tecnologia', `Immigrazione', `Salute', 'Infrastrutture e mobilità' and `Istituzioni e politica' (as reported in Table~\ref{tab:KSTopic}).
\begin{table}[ht]
    \centering
    \begin{tabular}{lrr}
\toprule
                    Topic &  KS Statistic &      p-value \\
\midrule
    Diritti della persona &      0.017 & 3.99e-01 \\
                 Ambiente &      0.010 & 9.93e-01 \\
        Economia e lavoro &      0.088 & 5.24e-09 \\
     Scienza e tecnologia &      0.068 & 1.32e-13 \\
             Immigrazione &      0.138 & 1.35e-04 \\
                   Salute &      0.079 & 1.80e-12 \\
Infrastrutture e mobilità &      0.123 & 4.83e-07 \\
   Istituzioni e politica &      0.149 & 9.07e-21 \\
    Sicurezza e giustizia &      0.0116 & 9.76e-01 \\
               Istruzione &      0.200 & 6.038e-01 \\
\bottomrule
\end{tabular}
    \caption{Kolmogorov-Smirnov (KS) statistic and p-value results for the comparison of overall rating distribution with each topic's rating distribution.}
    \label{tab:KSTopic}
\end{table}

\begin{figure}[ht]
    \centering
    \begin{subfigure}[b]{0.45\textwidth}
        \centering
        \includegraphics[width=\textwidth]{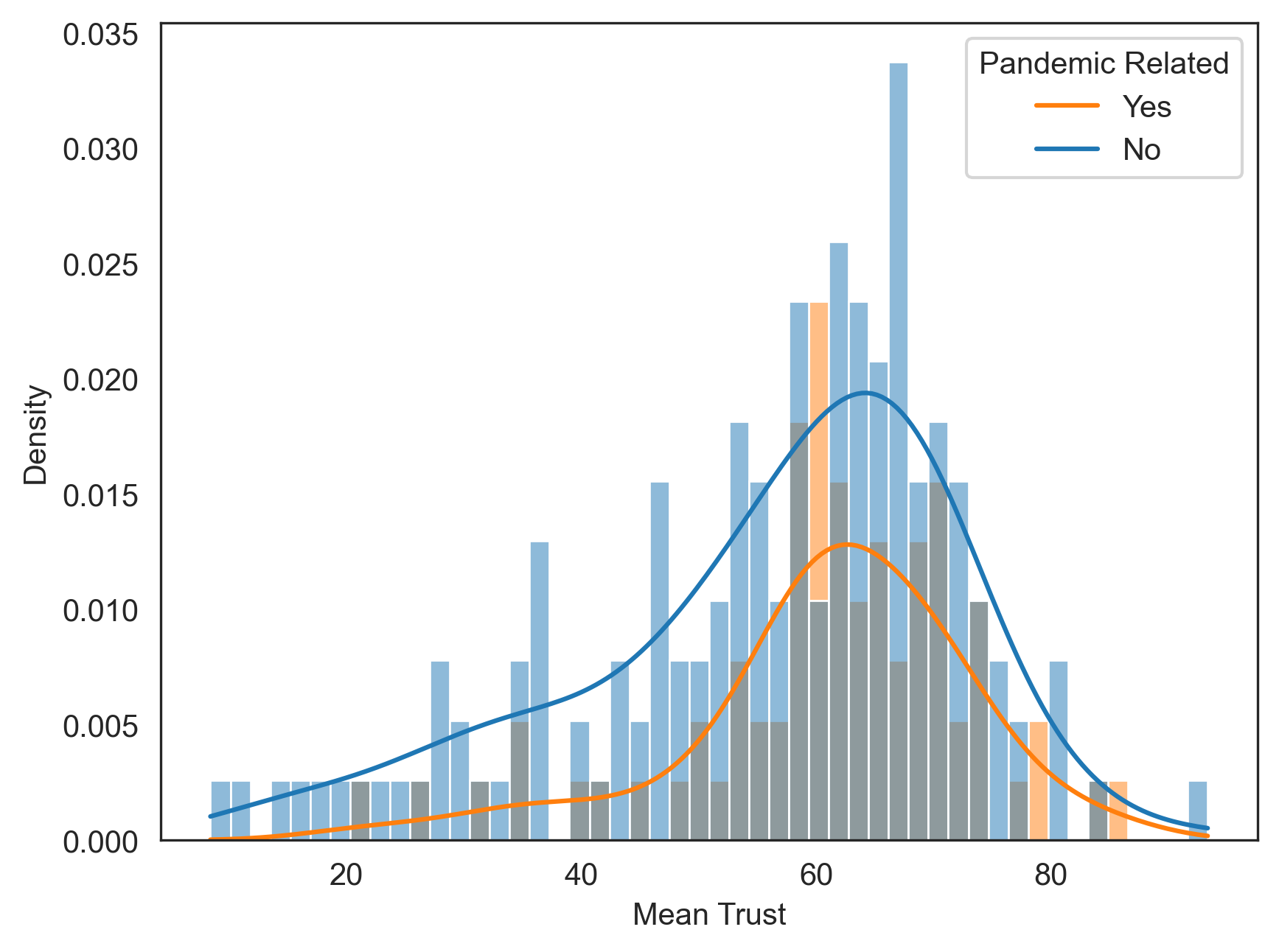}
        \caption{The panel displays the distribution of mean trust for news items related to the pandemic and those that are not.}
        \label{fig:trust-pandemic}
    \end{subfigure}
    \hfill
    \begin{subfigure}[b]{0.45\textwidth}
        \centering
        \includegraphics[width=\textwidth]{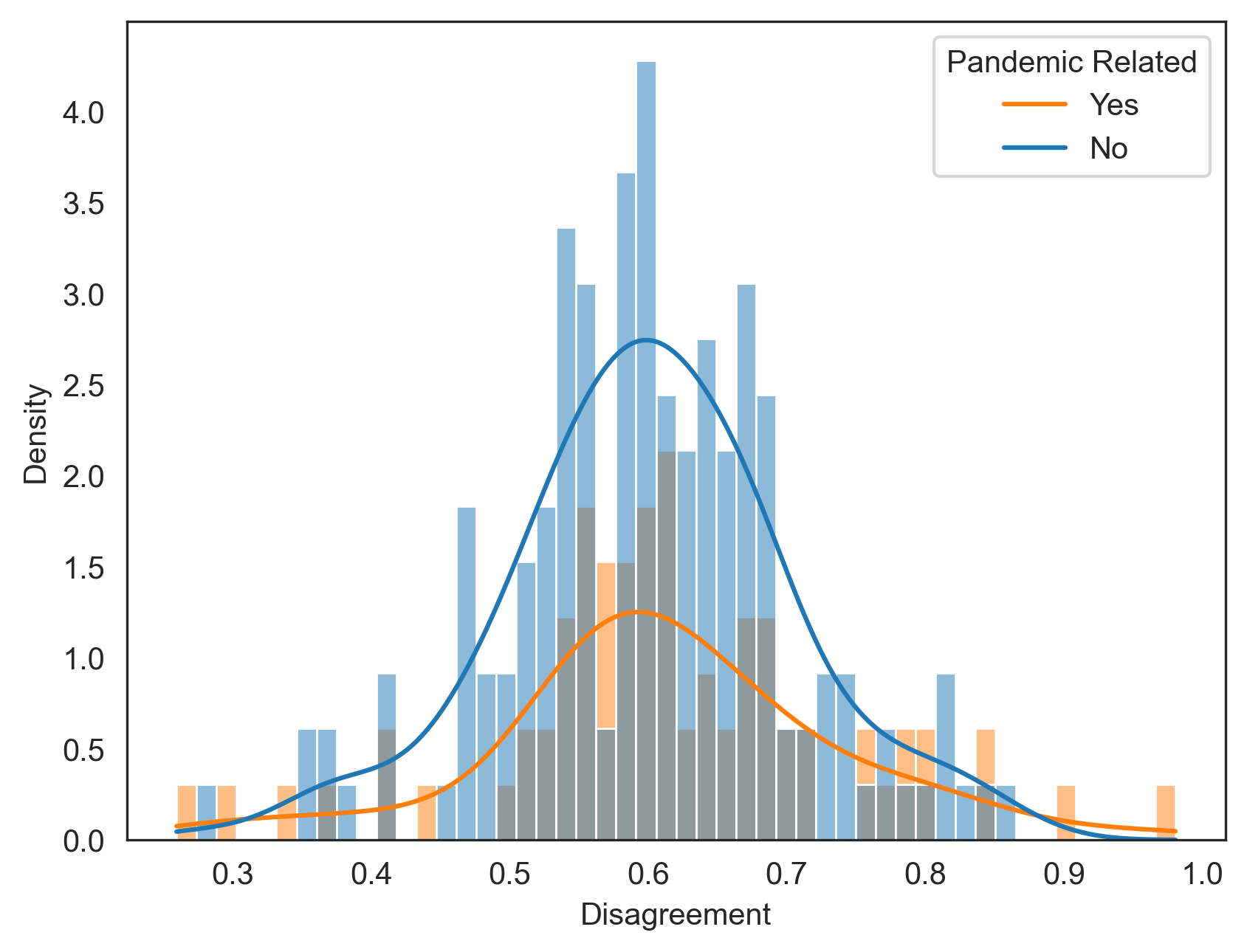}
        \caption{The panel displays the distribution of disagreement for news items related to the pandemic and those that are not.}
        \label{fig:disagreement-pandemic}
    \end{subfigure}
    \caption{The figures display the effect of the pandemic topic on the trust distribution.}
    \label{fig:figure_panel}
\end{figure}

Finally, the data includes a significant period that encompasses events related to the COVID-19 pandemic. As such, we test for any particular effect due to the presence of pandemic-related news items. In order to select news items related to the pandemic, we filter based on whether news items contain the following keywords: `covid', `coronavirus', `pandemic', `vaccine'. The words occur in the original dataset in 1210 articles. In the articles presented in the reviewed dataset they occur in 75 articles. We study the effect of the topic on the distribution of trust by evaluating changes in the mean  trust and in the disagreement scores (see Figure~\ref{fig:figure_panel}).

For mean trust we observe a significant shift in the distribution of mean trust (KS statistic 0.19 and p-value 0.037), pandemic related items have a higher mean trust over all, with the mean of pandemic related being 61, while the non pandemic related has a mean of 56. This could however been driven by a bias in the dataset that under-represents misinformation in the pandemic-related set of news items (see Table~\ref{tab:my_label}). Regarding disagreement, we do not observe any significant difference in the distribution related to the pandemic topic (KS statistic 0.09 and p-value 0.7).
\begin{table}[ht]
    \centering
    \begin{tabular}{l|rrr}    
    \toprule
    &  Misinformation   &    Yes &        No \\
        \midrule
    Pandemic-related &    &       &           \\

    False     &       &  0.51 &  0.16 \\
    True       &      &  0.31 &  0.02 \\
\bottomrule
\end{tabular}
    \caption{The fraction of each class of news item across Misinformation and Pandemic-related categories.}
    \label{tab:my_label}
\end{table}

\subsection{Modellisation of Trust}
To further investigate the effects of potential confounders on the distribution of trust we developed a regression model.
We define the model as follows:
\begin{equation}
  T_{ik}^{(t+1)} = \beta_0 + \beta_1 M_{i}^{(t+1)} + \beta_2 M_{i}^{(t)} + \beta_3 M_{i}^{(t-1)} + \vec{\theta} \vec{X}_i + \vec{\gamma} \vec{X}_{k}
\end{equation}

where:
\begin{itemize}
    \item $t \in \{1,2,3,...\}$ is the individual time of each participant as he participates in the ``Cartesio'' experience.
    \item $T_{ik}^{(t)}$ is the trust score assigned by individual $i$ to the news article $k$ rated at time $t$.
    \item $M_i^{(t)}$ is a binary value that is 1 participant $i$ rated a news item labeled as misinformation by the experts at time $t$.
    \item $\vec{X}_k$ is a vector of news characteristics assigned to news item $k$.
    \item $\vec{X}_i$ is a vector of characteristics assigned to participant $i$.
\end{itemize}

\begin{figure}
    \centering
    \includegraphics[width=\linewidth]{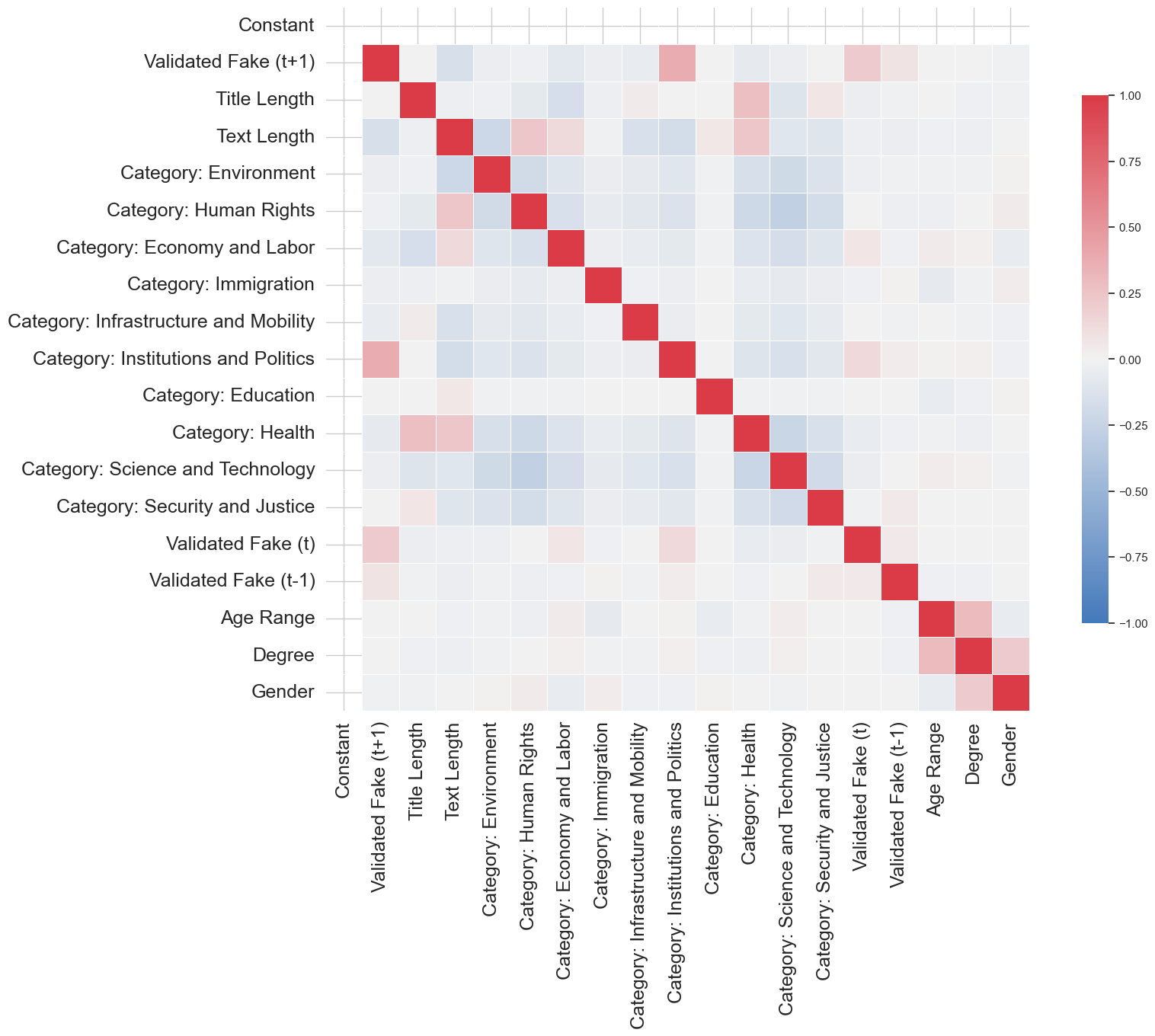}
    \caption{This heatmap visualises the correlation matrix of independent variables of the regression analysis. Positive correlations are indicated by red, and negative correlations by blue.}
    \label{fig:corpolot}
\end{figure}

We first perform a co-linearity investigation by checking correlation between the dependent variables of the model. The results can be seen in Figure~\ref{fig:corpolot}. We observe generally weak correlations outside of the age, degree, gender triangle.
We also compute the variance inflation factor for the non categorical variables and obtain a maximal value of 1.30 for title length, well within the bounds of acceptability of an ordinary least squares model.

We then perform two regressions, one with the raw data from the experiment and the other sample-weighted in order to correct for demographic imbalances found between the observed population and the Italian demographics.

\begin{figure}
    \centering   
    \begin{subfigure}[b]{0.8\textwidth}
        \centering
        \includegraphics[width=\textwidth]{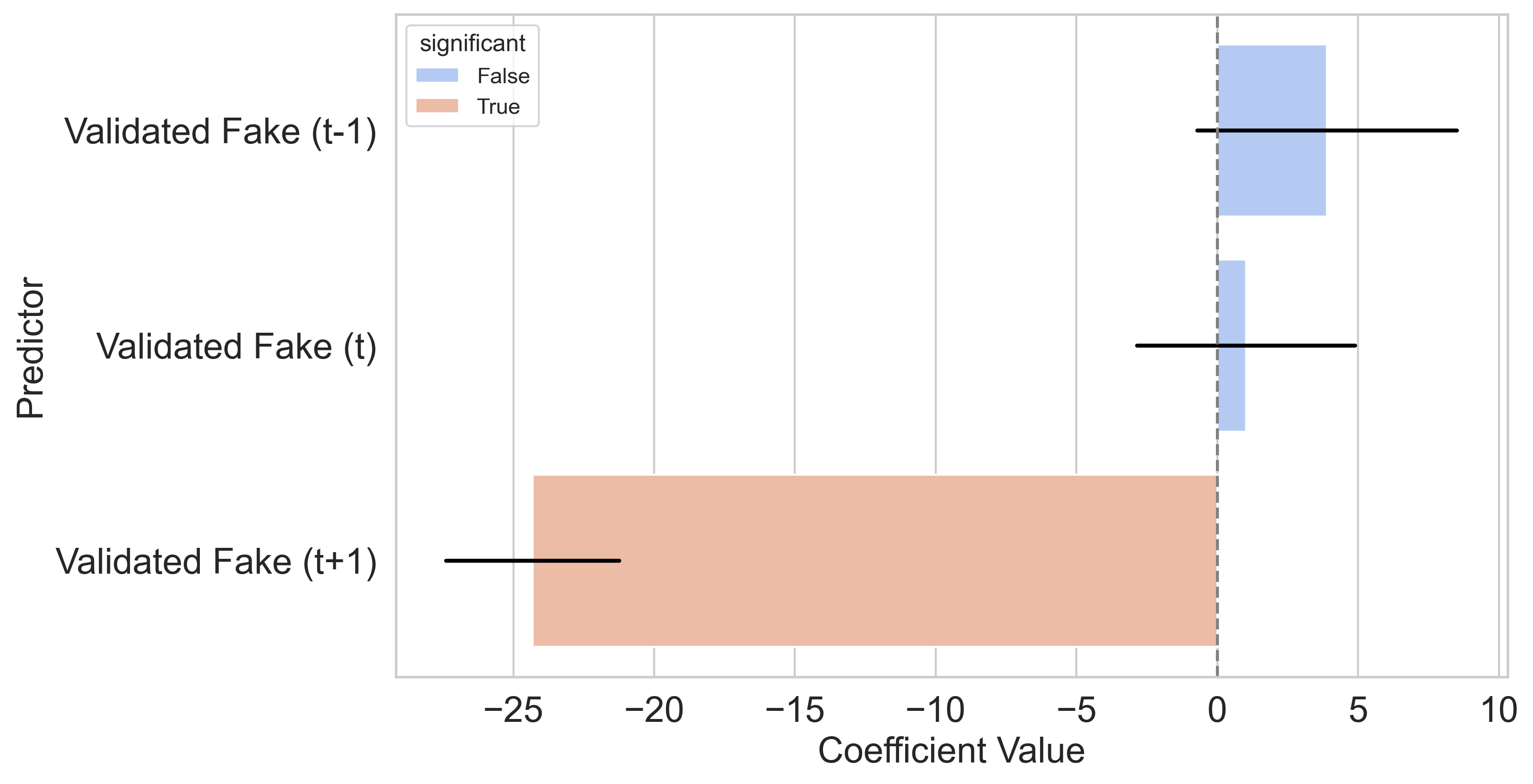}
        \caption{Misinformation Effects - Ordinary Regression}
    \end{subfigure}
    \vskip\baselineskip
    \begin{subfigure}[b]{0.8\textwidth}
        \centering
        \includegraphics[width=\textwidth]{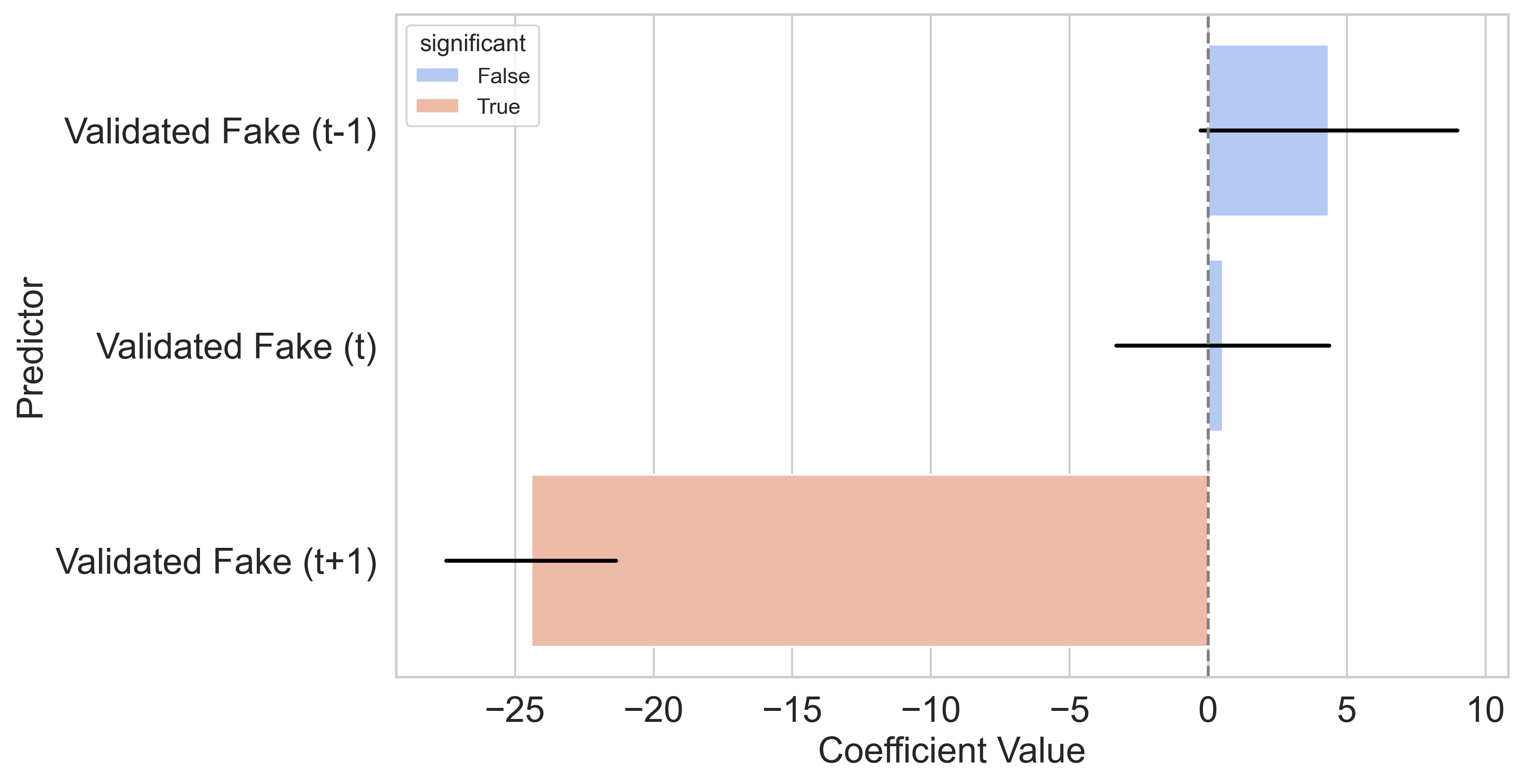}
        \caption{Misinformation Effects - Sample-Weighted Regression}
    \end{subfigure}
    \caption{Ordinary least-squares regression analysis results the misinformation at different times set of variables.}
    \label{fig:reg_mis}
\end{figure}

\begin{figure}
    \begin{subfigure}[b]{\textwidth}
        \centering
        \includegraphics[width=\textwidth]{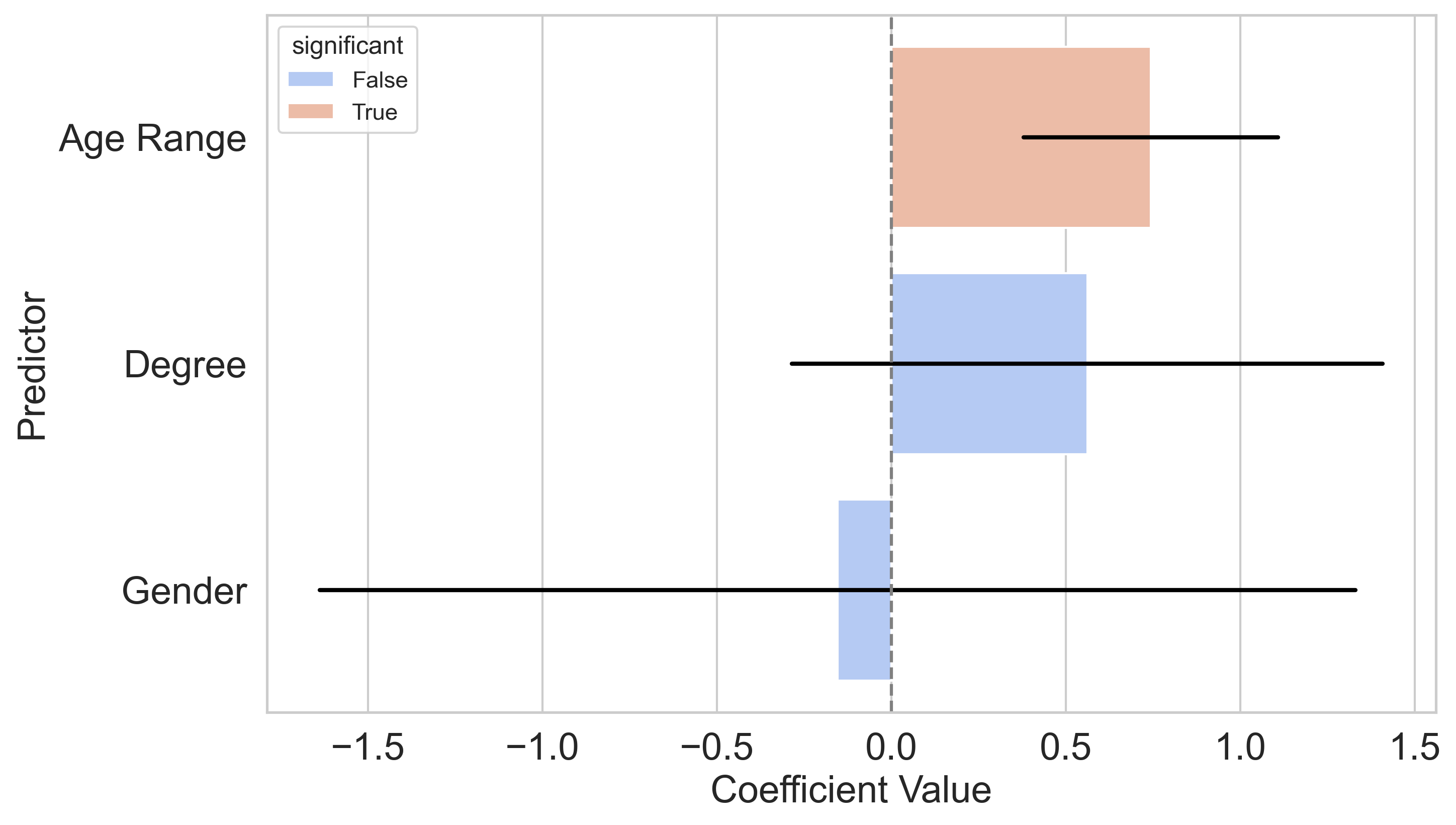}
        \caption{Demographics - Ordinary Regression}
    \end{subfigure}
    \vskip\baselineskip
    \begin{subfigure}[b]{\textwidth}
        \centering
        \includegraphics[width=\textwidth]{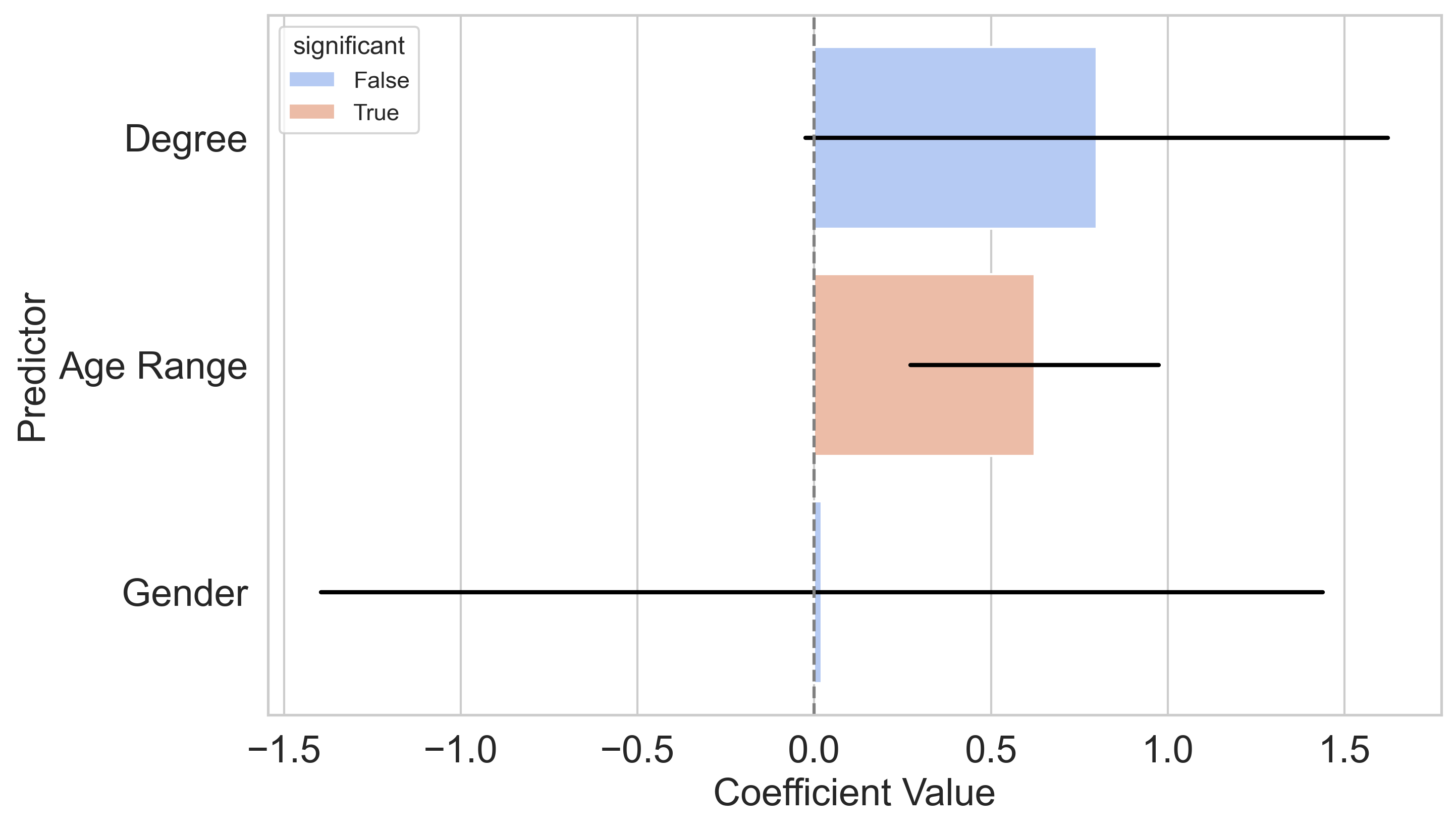}
        \caption{Demographics  - Sample-Weighted Regression}
    \end{subfigure}
    \caption{Ordinary least-squares regression analysis results the demographic properties set of variables.}
    \label{fig:reg_demo}
\end{figure}

\begin{figure}
    \begin{subfigure}[b]{\textwidth}
        \centering
        \includegraphics[width=\textwidth]{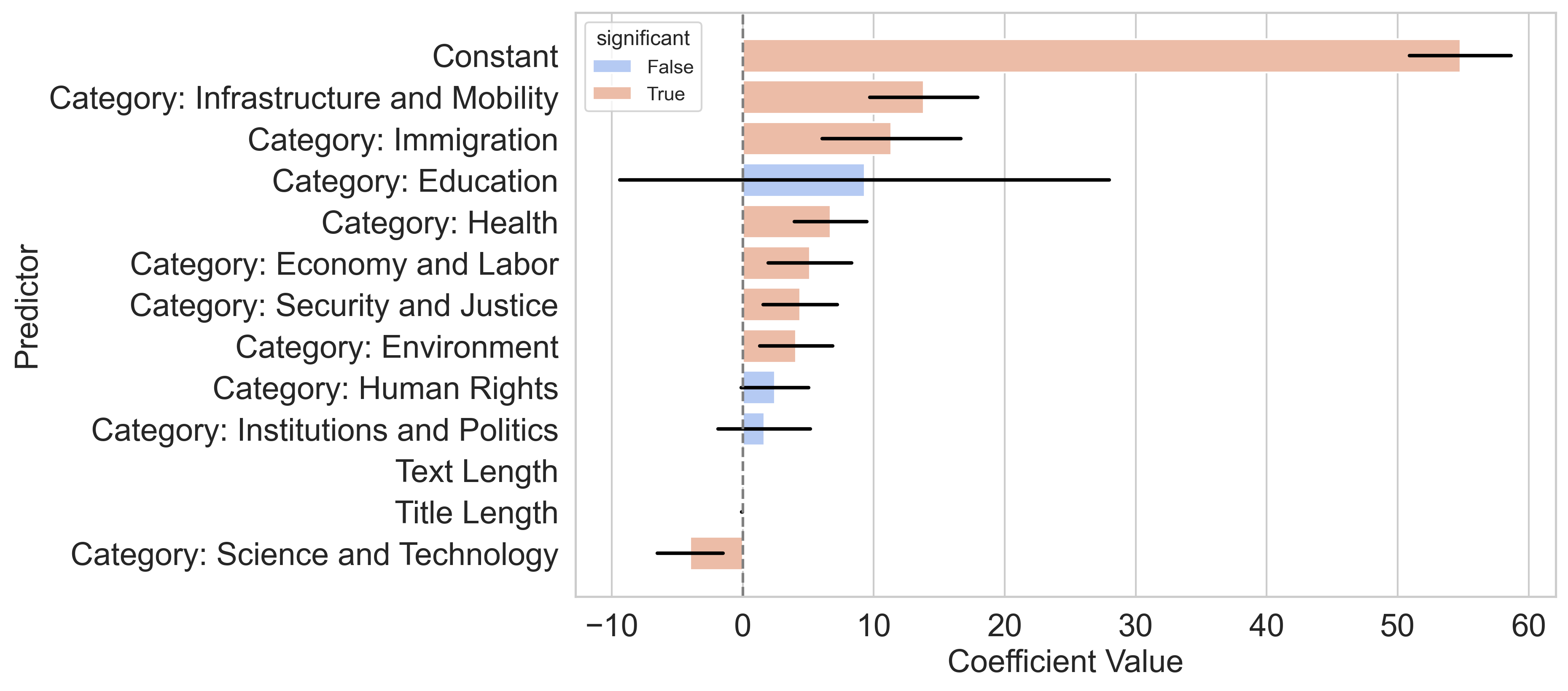}
        \caption{News Properties - Ordinary Regression}
    \end{subfigure}
    \vskip\baselineskip
    \begin{subfigure}[b]{\textwidth}
        \centering
        \includegraphics[width=\textwidth]{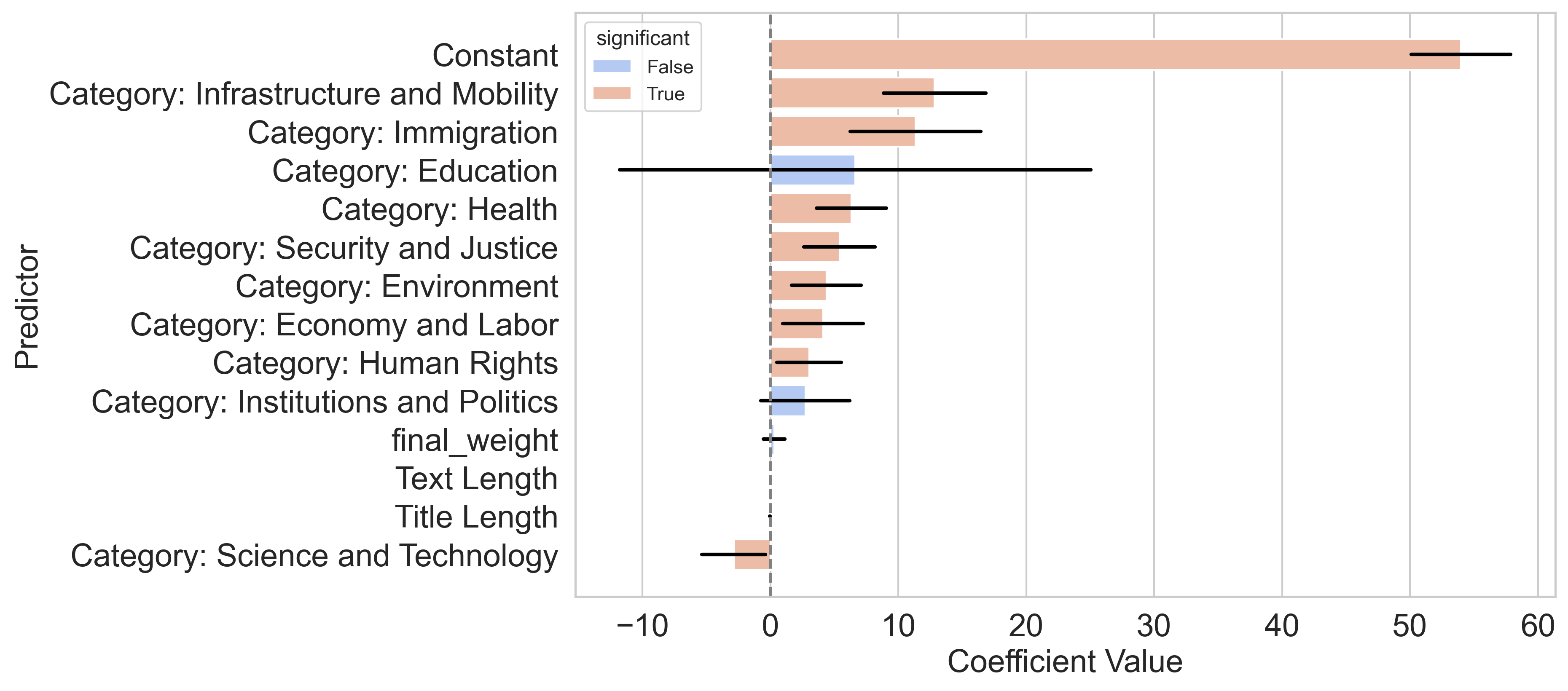}
        \caption{News Properties - Sample-Weighted Regression}
    \end{subfigure}

    \caption{Ordinary least-squares regression analysis results the news properties set of variables.}
    \label{fig:reg_news}
\end{figure}

In Figures~\ref{fig:reg_mis}, \ref{fig:reg_demo} and~\ref{fig:reg_news} the results of the regressions are displayed. Each figure has two panels, the upper for the non-weghted regression and the second for the sample-weighted regression. In the figures the significance of each coefficient is displayed through the color of the bars, red being significant and blue being non-significant. Significance is estimated through bonferroni multiple hypothesis adjusted p-values with a threshold set at $p \leq 0,05$.

We observe that misinformation has the greatest predictive power, but doesn't have a significant priming effect (previous misinformation labels are not significantly predictive). We also observe that age has a significant positive effect on trust in both regression models, while the other demographic determinants do not.
Finally, we observe small positive effects of news categories, specifically infrastructure and mobility, immigration, health, environment, security and justice and economy and labour. 

\section{Facebook Metrics Analysis}
\begin{figure}[H]
    \centering
    \includegraphics[width=.8\linewidth]{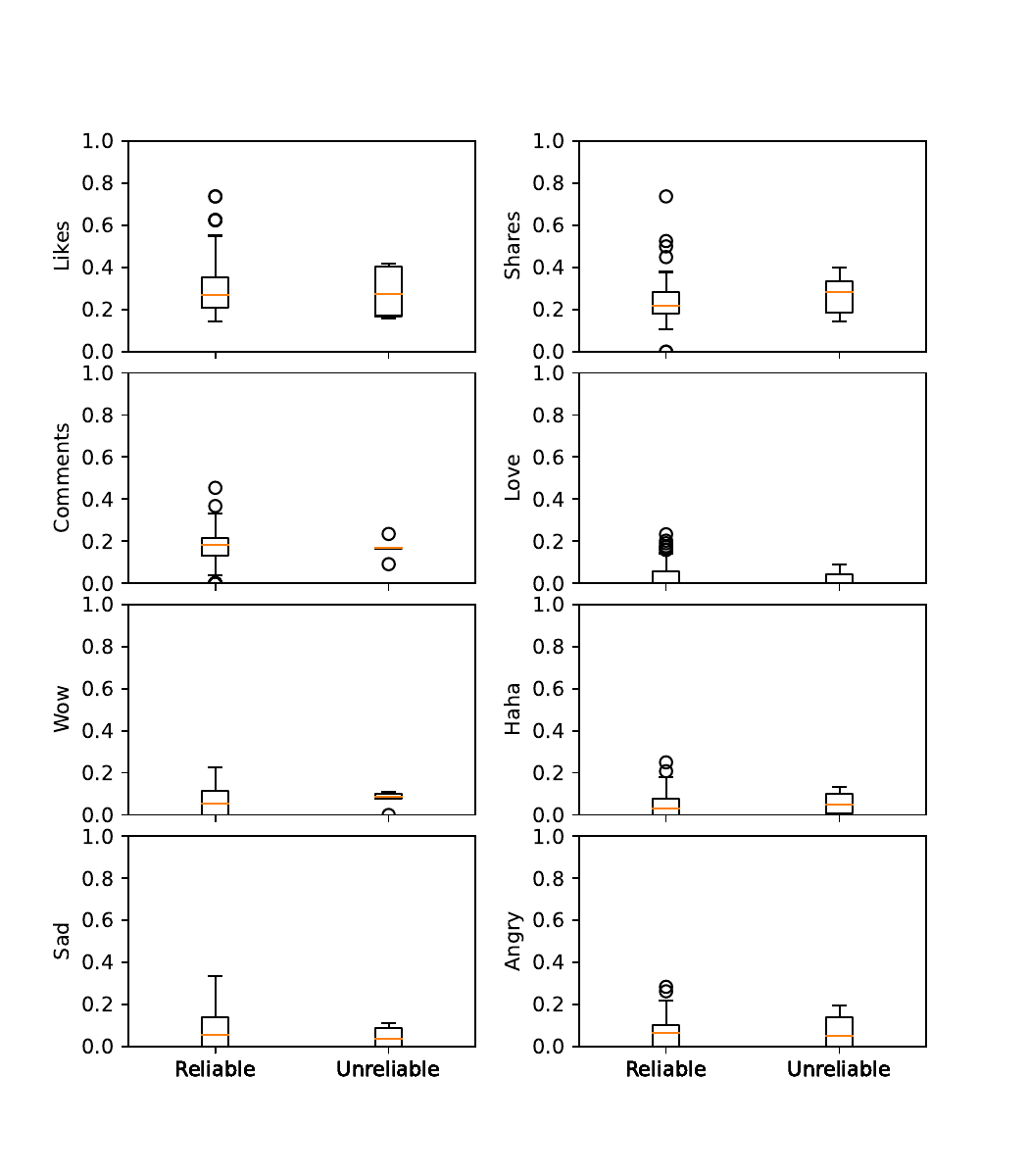}
    \caption{User engagement with reliable and unreliable content, across various reaction metrics on social media. The vertical axis indicates the proportion of total reactions (normalised scores from 0.0 to 1.0), while the horizontal axis categories the reactions by type: Likes, Shares, Comments, Love, Wow, Haha, Sad, and Angry. Each reaction type is split into two boxplots representing the distribution of reaction scores for reliable and unreliable content, respectively.}
    \label{fig:FBDistoFake}
\end{figure}

\begin{figure}[H]
    \centering
    \includegraphics[width=\linewidth]{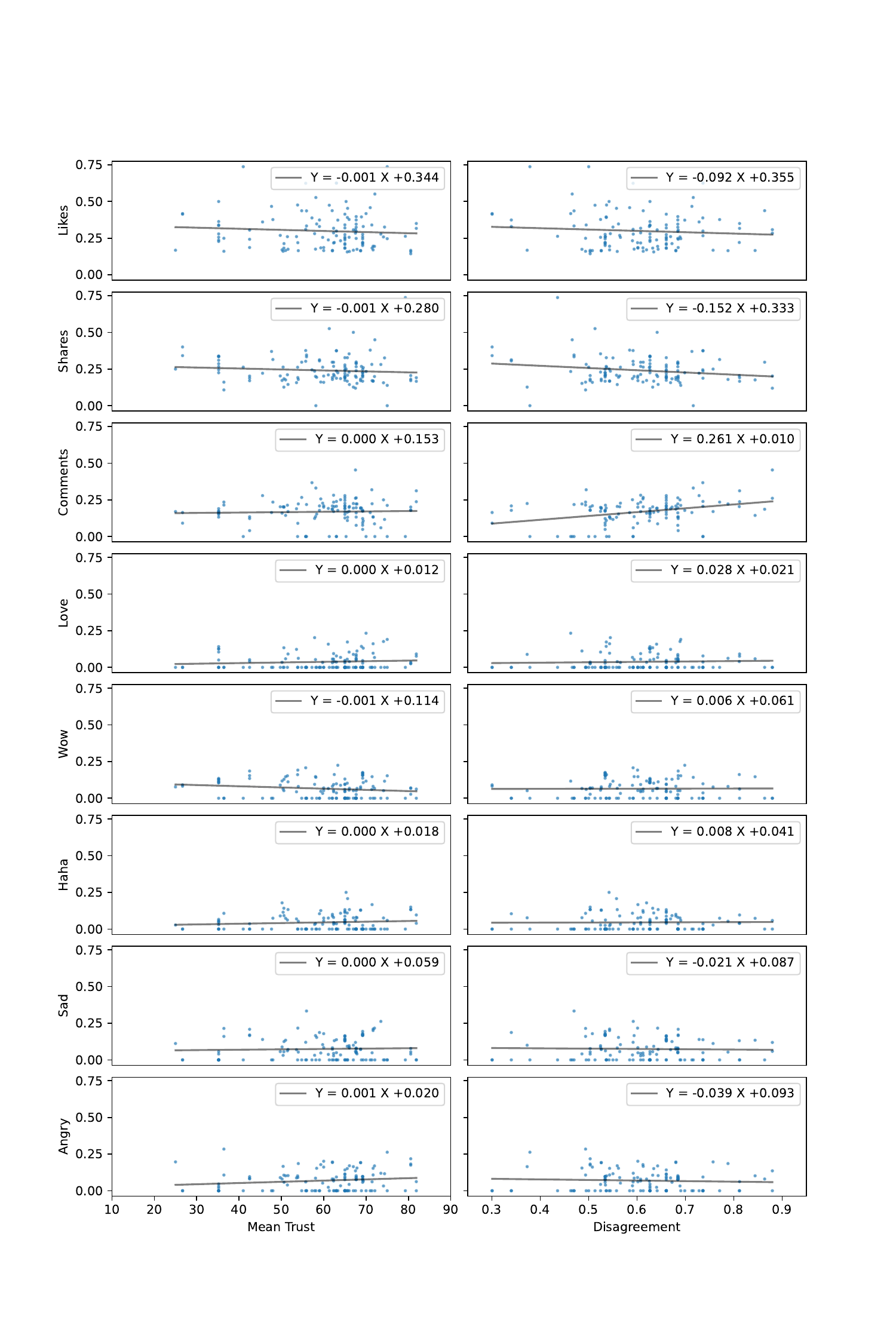}
    \caption{Scatter plots displaying the relationship between social media reactions (Likes, Shares, Comments, Love, Wow, Haha, Sad, Angry) and two independent variables: Mean Trust and Disagreement. The X-axis represents Mean Trust or Disagreement, and the Y-axis represents normalised scores of reactions (ranging from 0.00 to 0.75). Each plot includes a regression equation depicting the relationship between the social media reactions and the respective independent variable, highlighting how trustworthiness and disagreement levels influence user engagement metrics.}
    \label{fig:FBModels}
\end{figure}

\begin{figure}[H]
    \centering
    \includegraphics[width=0.8\linewidth]{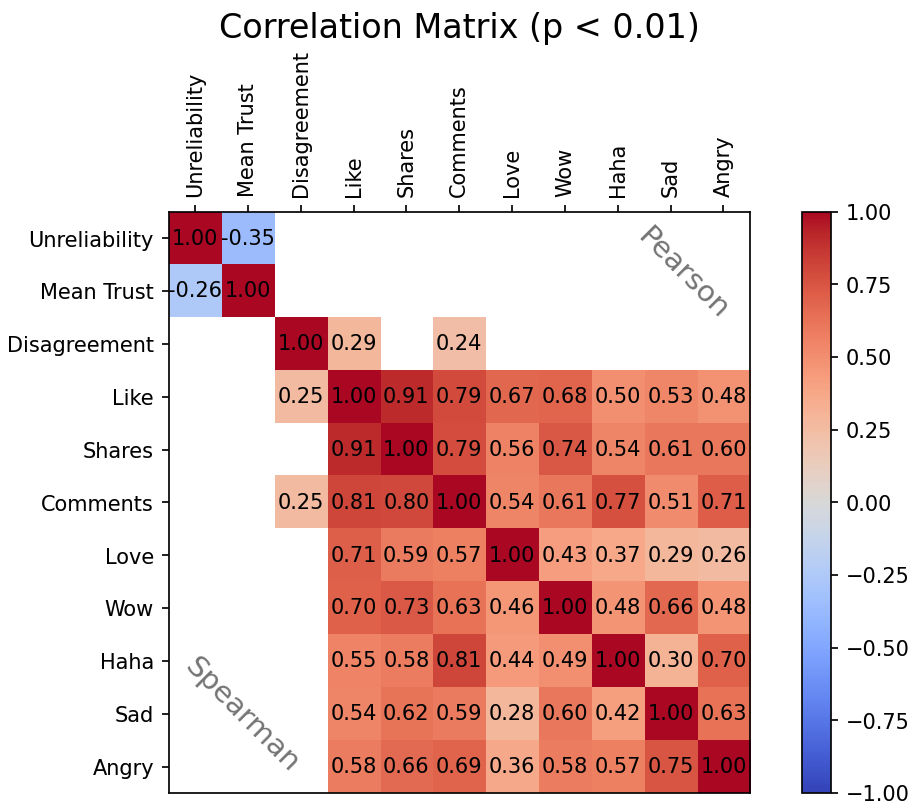}
    \caption{The correlation of Facebook engagement metrics with reliability and disagreement, before the popularity normalisation.}
    \label{fig:cornonorm}
\end{figure}

In order to better appreciate the effects of trust and misinformation on the metrics obtained when observing engagement of posts related to the news items we report the effect of expert given the misinformation label in Figure~\ref{fig:FBDistoFake} and the effect of disagreement and mean trust on Facebook measures in Figure~\ref{fig:FBModels}.
We also report the results of the correlation analysis before normalisation (Figure ~\ref{fig:cornonorm}). These results display are mainly driven by the effect of popularity, with all Facebook correlations being positive. 

\begin{longtable}{ll|ll}
\caption{The table displays the correlations shown in Figure 6 of the manuscripts. The number significance of the correlations are expressed using the `*' simbol as follows: `*' significance between $0.05$ and $10^{-4}$, `**' significance between $10^{-4}$ and $10^{-6}$ and `***' significance below $10^{-6}$} \\
\toprule
   variable 1 &      variable 2 & Pearson R & Spearman R \\
\midrule

\endfirsthead

\toprule
   variable 1 &      variable 2 & Pearson R & Spearman R \\
\midrule
\endhead
\midrule
\multicolumn{4}{r}{{Continued on next page}} \\
\midrule
\endfoot

\bottomrule
\endlastfoot
Unreliability &      Mean Trust &  -0.36 ** &    -0.26 * \\
Unreliability &    Disagreement &     -0.08 &       0.01 \\
Unreliability &            Like &     -0.03 &      -0.03 \\
Unreliability &          Shares &      0.07 &       0.07 \\
Unreliability &        Comments &     -0.01 &      -0.05 \\
Unreliability &            Love &     -0.06 &      -0.05 \\
Unreliability &             Wow &      0.05 &       0.07 \\
Unreliability &            Haha &      0.05 &       0.06 \\
Unreliability &             Sad &     -0.09 &      -0.08 \\
Unreliability &           Angry &       1.0 &       0.01 \\
Unreliability &    Entro. Diff. &      0.07 &       0.06 \\
Unreliability & Simm. K.L. Div. &    -0.2 * &    -0.23 * \\
   Mean Trust &    Disagreement &      0.01 &      -0.15 \\
   Mean Trust &            Like &     -0.08 &      -0.08 \\
   Mean Trust &          Shares &     -0.09 &      -0.16 \\
   Mean Trust &        Comments &      0.04 &       0.01 \\
   Mean Trust &            Love &       0.1 &       0.14 \\
   Mean Trust &             Wow &     -0.17 &      -0.12 \\
   Mean Trust &            Haha &      0.11 &        0.0 \\
   Mean Trust &             Sad &      0.04 &       0.05 \\
   Mean Trust &           Angry &      0.15 &       0.16 \\
   Mean Trust &    Entro. Diff. &      0.02 &       0.04 \\
   Mean Trust & Simm. K.L. Div. &     0.2 * &       0.16 \\
 Disagreement &            Like &     -0.08 &       0.02 \\
 Disagreement &          Shares &   -0.18 * &      -0.11 \\
 Disagreement &        Comments &   0.35 ** &     0.28 * \\
 Disagreement &            Love &      0.06 &       0.15 \\
 Disagreement &             Wow &      0.01 &      -0.01 \\
 Disagreement &            Haha &      0.02 &       0.07 \\
 Disagreement &             Sad &     -0.03 &      -0.03 \\
 Disagreement &           Angry &     -0.06 &      -0.06 \\
 Disagreement &    Entro. Diff. &    0.19 * &       0.14 \\
 Disagreement & Simm. K.L. Div. &   -0.22 * &    -0.22 * \\
         Like &          Shares &    0.4 ** &   0.66 *** \\
         Like &        Comments &  -0.38 ** &    -0.25 * \\
         Like &            Love &   -0.22 * &    -0.31 * \\
         Like &             Wow &   -0.4 ** &  -0.43 *** \\
         Like &            Haha &  -0.38 ** &  -0.55 *** \\
         Like &             Sad & -0.49 *** &  -0.58 *** \\
         Like &           Angry & -0.52 *** &  -0.71 *** \\
         Like &    Entro. Diff. & -0.68 *** &   -0.6 *** \\
         Like & Simm. K.L. Div. &     -0.09 &    -0.26 * \\
       Shares &        Comments &  -0.41 ** &   -0.37 ** \\
       Shares &            Love &   -0.22 * &     -0.3 * \\
       Shares &             Wow &    -0.3 * &    -0.31 * \\
       Shares &            Haha &  -0.36 ** &  -0.49 *** \\
       Shares &             Sad &   -0.32 * &  -0.43 *** \\
       Shares &           Angry & -0.51 *** &   -0.6 *** \\
       Shares &    Entro. Diff. &  -0.5 *** &  -0.47 *** \\
       Shares & Simm. K.L. Div. &     -0.11 &    -0.26 * \\
     Comments &            Love &       0.0 &       0.01 \\
     Comments &             Wow &     -0.12 &      -0.16 \\
     Comments &            Haha &    0.23 * &    0.39 ** \\
     Comments &             Sad &     -0.08 &      -0.04 \\
     Comments &           Angry &      0.03 &       0.04 \\
     Comments &    Entro. Diff. &   0.36 ** &     0.23 * \\
     Comments & Simm. K.L. Div. &    -0.2 * &      -0.07 \\
         Love &             Wow &      0.13 &     0.21 * \\
         Love &            Haha &      0.03 &     0.18 * \\
         Love &             Sad &     -0.15 &      -0.02 \\
         Love &           Angry &     -0.07 &       0.09 \\
         Love &    Entro. Diff. &      0.05 &       0.11 \\
         Love & Simm. K.L. Div. &      0.01 &      -0.04 \\
          Wow &            Haha &     -0.04 &       0.05 \\
          Wow &             Sad &     0.2 * &     0.23 * \\
          Wow &           Angry &      0.09 &     0.21 * \\
          Wow &    Entro. Diff. &  0.46 *** &   0.49 *** \\
          Wow & Simm. K.L. Div. &     -0.11 &      -0.09 \\
         Haha &             Sad &      -0.1 &       0.03 \\
         Haha &           Angry &    0.23 * &      0.3 * \\
         Haha &    Entro. Diff. &    0.28 * &     0.34 * \\
         Haha & Simm. K.L. Div. &      0.12 &       0.15 \\
          Sad &           Angry &    0.34 * &   0.48 *** \\
          Sad &    Entro. Diff. &  0.49 *** &   0.56 *** \\
          Sad & Simm. K.L. Div. &    0.18 * &       0.09 \\
        Angry &    Entro. Diff. &    0.27 * &    0.37 ** \\
        Angry & Simm. K.L. Div. &    0.34 * &     0.24 * \\
 Entro. Diff. & Simm. K.L. Div. &   -0.19 * &      -0.08 \\
 \label{correlations}
\end{longtable}
